\documentclass[11pt]{article}%
\usepackage{afterpage}
\usepackage{bm}
\usepackage{amsmath}
\usepackage{amsfonts}
\usepackage{tabularx}
\usepackage{amssymb}
\usepackage{booktabs}
\usepackage{float}
\newcolumntype{C}{>{\centering\arraybackslash}X}
\usepackage{graphicx}
\usepackage[subrefformat=parens]{subcaption}
\usepackage{color}
\usepackage{comment}
\usepackage{rotating}
\usepackage{xcolor}

\usepackage[height=9in,left=1in,right=0.75in,bottom=1in]{geometry}
\usepackage[breaklinks=true,bookmarksopen=true,colorlinks=true,citecolor=blue,urlcolor=blue]{hyperref}%

\usepackage[authoryear]{natbib}
\usepackage{booktabs}
\usepackage{pdfsync}
\usepackage{pdflscape}
\usepackage{multirow}
\usepackage{setspace}
 \usepackage{subcaption}
\usepackage{multicol}%
\providecommand{\U}[1]{\protect\rule{.1in}{.1in}}
\newtheorem{theorem}{Theorem}[section]

\newtheorem{assumption}{Assumption}

\newtheorem{lemma}[theorem]{Lemma}

\newtheorem{proposition}[theorem]{Proposition}
\newtheorem{remark}{Remark}[section]

\newenvironment{proof}[1][Proof]{\noindent\textbf{#1.} }{\ \rule{0.5em}{0.5em}}

\makeatletter

\long
\def\comment#1{}
\@addtoreset{lemma}{Appendix B} \makeatother

\begin{document}

\title{Regression Discontinuity Design with Multivalued Treatments\thanks{We would
like to thank numerous seminar participants at various departments and conferences since 2014 for
useful comments. First version: June 9, 2015, available at SSRN:
\href{https://ssrn.com/abstract=2615846}{https://ssrn.com/abstract=2615846}.}}
\author{Carolina Caetano\\\textit{University of Georgia}
\and Gregorio Caetano\\\textit{University of Georgia}
\and Juan Carlos Escanciano\thanks{Research funded by the Spanish
Programa de Generaci\'{o}n de Conocimiento, reference number
PGC2018-096732-B-I00. }\\\textit{Universidad Carlos III de Madrid}}
\date{June 2020}
\maketitle

\begin{abstract}
We study identification and estimation in the Regression Discontinuity Design
(RDD) with a multivalued treatment variable. We also allow for the inclusion of covariates. We show that without
additional information, treatment effects are not identified. We give necessary and sufficient conditions  that lead to identification of LATEs as well as of weighted averages of the conditional  LATEs. We show that if the first stage discontinuities of the multiple treatments conditional on covariates are linearly independent, then it is possible to identify multivariate weighted averages of the treatment effects with convenient identifiable weights. If, moreover, treatment effects do not vary with some covariates or a flexible parametric structure can be assumed, it is possible to identify (in fact, over-identify) all the treatment effects. The over-identification can be used to test these assumptions. We propose a simple estimator, which can be programmed in
packaged software as a Two-Stage Least Squares regression, and packaged standard errors and tests can also be used. 
Finally, we implement our approach to identify  the effects of
different types of insurance coverage on health care utilization, as in
\citet{CardMedicare}.

\begin{description}
\item[Keywords:] Regression Discontinuity Design; Multiple Treatments; Over-identifying restrictions; Nonparametric; LATE; Medicare; Health Insurance.

\item[\emph{JEL classification:}] C13; C14; C21; I13

\end{description}
\end{abstract}


\newpage
\section{Introduction}

Regression Discontinuity Design (RDD) has emerged as one of the most credible
identification strategies in the social sciences; see \cite{Imbens_Lemieux}
and \cite{Lee_Lemieux} for early surveys of the literature and
\cite{Cattaneo_Escanciano} and \cite{Cattaneo_Idrobo_Titiunik} for more recent overviews. The vast majority of
research on RDD focuses on the binary treatment case. In this paper
we study RDD with a multivalued treatment.\footnote{Not to be confused with
having multiple running variables or multiple thresholds, for which several
proposals are available; see the review in \cite{Cattaneo_Escanciano}.} 

The motivation to study this case is the prevalence of empirically relevant
situations in economics in which an RDD approach is undertaken and the treatment variable
is multivalued. See, for example, empirical applications in \cite{AngristLavy99},
\cite{ChayG}, \cite{LudwigMiller}, \cite{Carpenter_Dobkin}, \cite{McCrary_Royer}, \cite{Brolloetal}, \cite{Pop-Eleches_Urquiola},  \cite{Buser}, 
\cite{Isen_Rossin-Slater_Walker}, \cite{Spenkuch_Toniatti}, \cite{Nakamura_Steinsson}, \cite{Campante_Yanagizawa-Drott}, \cite{Corbi_Papaioannou_Surico},
\cite{Agarwal_Chomsisengphet_Mahoney_Stroebe}, \cite{Dell_Querubin} \cite{Finkelstein_Hendren_Shepard}, \cite{Dube_Giuliano_Leonard}, and \cite{Fort_Ichino_Zanella}, among
others. In such cases, the RDD estimate consists of the ratio of the discontinuity of the average outcome at the threshold divided by the corresponding discontinuity of the average treatment. This quantity is hard to interpret unless further assumptions are made. For example, if the effect of the treatment is equal across different treatment intensities, then the RDD estimates the local average treatment effect (LATE). If the constant marginal effects assumption is too strong and we would like to explore different marginal effects for different values of the treatment variable, currently there exists  no method that can guide such undertaking. 

In this paper we examine the problem of identification of the effects of multiple  treatments in the RDD setting with a single running variable and threshold. We also allow for the inclusion of covariates.   We show that in the multivalued treatment case the LATEs are generally not identifiable under the standard RDD assumptions. A less ambitious goal is to identify averages  of the marginal effects at a given treatment value while avoiding contamination from the marginal treatment effects at other treatment values. We show that this is generally impossible. However, it is possible to avoid this type of contamination in expectation. The idea is to weight the different marginal effects in such a way that the weights of the marginal effect in which we are interested average to one, and the weights of the other marginal effects average to zero.

We show that in order to achieve ``separation in expectation" for the marginal effects of all treatment values, it is necessary and sufficient that the vectors of first stage discontinuities of all different treatments conditional on covariates are linearly independent. Specifically, we need as many linearly independent vectors of first stage discontinuities as there are treatment values. This condition is testable, and in fact it usually  holds, as long as the data has enough covariates. 

In order to identify all LATEs, as opposed to only separate marginal effects inside of an expectation, it is necessary to reduce the parameter space. We propose two strategies, the first supposes homogeneity of the LATEs in some covariates, the second supposes that the LATEs satisfy a flexible parametric model of the covariates (e.g. linear). Both strategies  often lead to over-identification of the LATEs, and hence, testability of the assumptions. 

Based on our identification strategy we propose an estimator which can be programmed as a weighted Two Stage Least Squares regression using packaged software. The packaged standard errors obtained from this regression can be used for inference. Packaged $t,$ $F$ and over-identification tests can be used to test the identification assumptions.  

Our paper relates to a number of studies discussing the inclusion of
covariates in the standard RDD setting; see, e.g., \cite{Imbens_Lemieux},
\cite{Frolich}, \cite{Calonico_Cattaneo_Farrell_Titiunik} and \cite{Frolich_Huber}, all of which deal with the binary case. The specific way in which we include the covariates is new. In the standard RDD setting covariates have been used for efficiency purposes, but we show that in the multivalued treatment setting covariates have the potential to help with the identification of multiple effects. Also related is the recent proposal by \cite{Dong_Lee_Gou}, which uses a control function approach with a scalar unobservable first stage heterogeneity to identify heterogeneous effects with a continuous scalar treatment.

We apply our approach to the problem of estimating the effects of Medicare insurance
coverage on health care utilization with a regression discontinuity design, as
in \citet{CardMedicare}. They exploit the fact that Medicare eligibility
varies discontinuously at age 65. Medicare eligibility may affect health care
utilization via two channels: (1) the extensive margin, because Medicare
eligibility provides coverage to people who were previously uninsured, and (2)
the intensive margin, because it provides more generous coverage to people who
were previously insured by other insurance policies. We find
that minorities and people with less education are more likely to be affected
by Medicare eligibility in the extensive margin (i.e. one insurance vs. no
insurance), while whites and people with higher education are more likely to
be affected by Medicare eligibility in the intensive margin, (i.e. more
generous vs. less generous insurance). Using our approach to exploit this
heterogeneity in the first stages allows us to identify the partial effects at
both these margins under testable assumptions. While the extensive margin seems to matter for recurrent,
lower cost, health care utilization (e.g., doctor visits), the intensive margin seems
to matter for sporadic, higher cost, health care utilization (e.g., hospital visits).

The paper is organized as follows. Section \ref{identification} presents the multivalued RDD model and studies functions of treatment effects which can be identified with parsimonious assumptions. Section \ref{sec:exclusion} discusses the further assumptions needed if one wants to identify the specific LATEs.  Section \ref{RDDestimation}
presents the estimators and their asymptotic behavior.  Section \ref{application2} presents the
application of our method to the problem of identifying the effects of
different types of insurance coverage on health care utilization. Finally, we
conclude in Section \ref{conclusion}. An Appendix contains proofs of the identification and asymptotic results.

\section{Identification of Treatment Effects\label{identification}}

\subsection{Model Setup}
We consider a setting of potential outcomes with a multivalued treatment
variable. Let $T_{i}$ be a treatment variable, with discrete support
$\mathcal{T}=\{t_{0},...,t_{d}\},$ $1\leq d<\infty,$ and where $t_{0}<t_{1}%
<\cdots<t_{d}.$ Let $Y_{i}(j)$ denote a potential outcome under treatment
level $T_{i}=t_{j}.$ Define $\alpha_{i}=Y_{i}(0),$ $\beta_{ij}=Y_{i}%
(j)-Y_{i}(j-1)$ and the $d$-dimensional vectors $\beta_{i}=(\beta
_{i1},...,\beta_{id})^{\prime}$ and $X_{i}=(1(T_{i}\geq t_{1}),...,1(T_{i}\geq
t_{d}))^{\prime},$ where $1(A)$ denotes the indicator function of the event
$A$ and $B^{\prime}$ denotes the transpose of the matrix or vector $B.$ Then, note that the observed
outcome $Y_{i}=Y_{i}(T_{i})$ can be expressed as the random coefficient model %
\begin{equation}
Y_{i}=\alpha_{i}+\beta_{i}^{\prime}X_{i}.\label{RC}%
\end{equation}
Self-selection into treatment makes the vector of treatment indicators $X_{i}$
potentially correlated with the vector of treatment effects $\beta_{i}.$ We consider an extension of the RDD
identification strategy in \cite{HTV2001}. The extension is along two
dimensions: (i) a multivariate endogenous variable and (ii) allowing for the presence of
covariates $W_{i}.$ Let $\mathcal{W}$
denote the support of the distribution of $W_{i}.$  Henceforth, we assume that the running variable $Z_{i}$ is
univariate and continuously distributed, and the RDD threshold is $z_0.$ Denote by $T_{i}(z)$ the potential
treatment variable for someone with $Z_{i}=z$ and $X_{ij}(z)=1(T_{i}(z)\geq
t_{j})$ the corresponding potential treatment indicator.

To aid in the understanding of our results, we introduce here the example discussed in our application (see \citet{CardMedicare} and Section \ref{application2}.) Consider the problem of estimating the effect of having Medicare health insurance on whether a person utilizes health care services. Here we would like to allow the marginal effect of going from no insurance to one insurance (Medicare) to be different from the marginal effect of going from one insurance to two or more insurances (Medicare plus additional insurances). Having another insurance in addition to Medicare can be beneficial because the other insurance may pickup Medicare's copays, or have maximum out-of-pocket  limits, or be more generous for some specific health care events. Thus, in this example $X_{i1}$ is an indicator that $i$ holds at least one insurance policy, and $X_{i2}$ is an indicator that $i$ holds at least two insurance policies. The running variable $Z_i$ represents $i$'s age (measured in quarter increments), and the threshold $z_0=65$ represents eligibility to Medicare.  Our data also has information on the person's race (white, denoted WH or minority, denoted MIN) and education (high school dropout or less, denoted DRP, high school graduate but no more, denoted HS, and more than high school, such as at least some college or other higher education, denoted COL), which will be useful.\footnote{Our data also has information on gender, Hispanic status, region and year of the sample, which we use in our application in Section \ref{application}. For instructional purposes we explore only race and education in the theory illustrations.}

The following assumption generalizes the conditions of the standard RDD to allow for a multivalued treatment and the presence of covariates. Define
$\Delta_{ij}(e)=X_{ij}(z_{0}+e)-X_{ij}(z_{0}-e)$ and the conditional moment%
\[
\beta_{j}(w,e)=E[\beta_{ij}|W_{i}=w,\Delta_{ij}(e)=1].
\]

\begin{assumption}
\label{assumption structural RDD} (i) $E[Y_{i}(0)|Z_{i}=z,W_{i}=w]$ is
continuous in $z$ at $z_{0}$ almost surely (a.s.) in $w\in\mathcal{W}$; for
each $j:$ (ii) $(\beta_{i},X_{ij}(z))$ is independent of $Z_{i}$ near $z_{0}$
conditionally on $W_{i}$; (iii) there exists $\varepsilon>0$ such that for all
$0<e<\varepsilon:$ $\Delta_{ij}(e)\geq0$ a.s. conditionally on $W_{i};$
and (iv) $\lim_{e\downarrow0}\beta_{j}(w,e)$ exists and $\left\vert \beta
_{j}(w,e)\right\vert \leq g_{j}(w)$ with $E[g_{j}(W)]<\infty.$
\end{assumption}

In our application, Assumption \ref{assumption structural RDD} requires the following: conditional on $W$ (which could be a constant, race, education, or both), (i) the probability that a person without insurance will utilize health care services is continuous at the Medicare eligibility threshold, $z_0=65$; (ii) the  effects of going from zero to one, and from one to two or more insurances do not depend on age close to the threshold. Additionally, the likelihood that someone would go from zero to one insurance or from one to two or more insurances is not affected by age (except insofar as it allows the person to qualify to Medicare) close to the threshold;  (iii) a person slightly older than 65 must hold as many policies or more than they would have held if they were slightly younger than  65 (this is a local monotonicity condition); and  (iv) the  effects of going from zero to one, and from one to two or more insurances for people slightly older than 65 are bounded and continuous in age at 65.

Assumption \ref{assumption structural RDD}(iv) guarantees that the following limit
is well defined,
\begin{equation*}
\beta_{j}(w)=\lim_{e\downarrow0}E[\beta_{ij}|W_{i}=w,\Delta_{ij}(e)=1].
    \end{equation*}
The parameter $\beta_{j}(w)$ represents the conditional Local Average Treatment Effect (LATE)
 of moving from treatment level $t_{j-1}$ to $t_{j}$ at $Z_{i}=z_{0}$
for someone with $W_{i}=w.$ In other words,  the conditional LATEs $\beta_j(w)$ are averages that are taken conditional on $W=w,$ and locally both at the threshold point and among compliers (compliers for variable $X_j$ in this context are defined as observations such that $\lim_{e\rightarrow 0}\Delta_{ij}(e)=1$). In our application $\beta_{1}(W)$ represents the expected 
effect  of going from no insurance to one insurance at the age of $65,$ among compliers,  conditional on $W,$ and  $\beta_2(W)$ is analogously defined. The parameter of interest is therefore the vector $\beta(w)=(\beta
_{1}(w),...,\beta_{d}(w))^{\prime}.$

A less ambitious quantity is the unconditional vector of LATEs $\beta
=(\beta_{1},\dots,\beta_{d})^{\prime},$ where
\[
\beta_{j}=\lim_{e\downarrow0}E[\beta_{ij}|\Delta_{ij}(e)=1].
\]
Our results always include this as a special case, and we refer to this as the
\textquotedblleft$W$ is constant" case, since conditioning on a constant is
equivalent to not conditioning on it. Note also that the LATE of $X_{j}$,
$\beta_{j}$, can be obtained from the conditional LATEs of $X_{j}$, $\beta
_{j}(W)$, since under our conditions,
\begin{equation}
\beta_{j}=\lim_{e\downarrow0}E\left[  \beta_{j}(W)\big|\Delta_{ij}%
(e)=1\right]  ,\label{eq:media}%
\end{equation}
by the law of iterated expectations and  Assumption \ref{assumption structural RDD}%
(iv). Then, whenever the $\beta_{j}(W)$ are
identified a.s., so are the $\beta_{j}.$

Assumption \ref{assumption structural RDD} and equation (\ref{RC}) yield the
key identifying equation
\begin{equation}
\delta_{Y}(w)=\beta(w)^{\prime}\delta_{X}(w),\quad\text{for each }%
w\in\mathcal{W}, \label{rddlinear}%
\end{equation}
where, for a generic random vector $V,$ we use the notation
\[
\delta_{V}(w)=\lim_{z\downarrow z_{0}}E[V|Z=z,W=w]-\lim_{z\uparrow z_{0}%
}E[V|Z=z,W=w].
\]
When $W$ is constant, we denote the unconditional discontinuities $\delta
_{V}.$ Equation (\ref{rddlinear}) relates first stages $\delta_{X}(w)$ and
reduced form effects $\delta_{Y}(w)$  with the structural parameters of interest $\beta(w)$.

\subsection{Can Multivariate Local Average Treatment Effects be identified?}\label{sec:impossibility}
In this section we study the identifiability of the $\beta_j(W)$ (including the LATE $\beta_j,$ when $W$ is constant) and show that identification is generally impossible. Nevertheless, the propositions in this section allow us to learn what is necessary for identification and which parsimonious conditions may be brought in to improve identification, which is what we do in the following sections. All proofs are gathered in the Appendix.

\begin{proposition}
\label{Lack}Under Assumption \ref{assumption structural RDD}, $\beta(w)$ is
not identifiable.
\end{proposition}

 The reason for lack of identification is intuitive: there is one equation  (\ref{rddlinear}) and $d$ coefficients to identify for each value of $w.$ Even if $W$ assumes many values, and thus we have many equations \eqref{rddlinear}, the number of coefficients to identify increases by $d$ with each new equation. The best we can identify is a linear combination of coefficients, and thus there is only partial identification. This impossibility result
implies that additional restrictions are necessary in order  to identify
meaningful causal parameters of interest.

This impossibility is  not due to the inclusion of the covariates, it is due to the multiple treatments. In particular, if $W$ is constant, this theorem implies the impossibility of identification of the unconditional LATEs $\beta_j$ unless further assumptions are made. 

One setting where identification of a causal parameter holds is when, for some group characterized by a particular covariate value $w,$ all but one
first stage coefficients are zero. The following proposition states that this case is, in fact, the only circumstance in which identification of LATEs can be achieved, unless further assumptions are made. For any vector $a,$ let $a_{-j}$ denote $a$ without
the $j$-th coordinate.

\begin{proposition}
\label{Local}Under Assumption \ref{assumption structural RDD}, $\beta_{j}(w)$ is identified $\iff$  $\delta_{X_{-j}}(w)=0$ and $\delta_{X_{j}}(w)\neq0.$ In this case, 
$\beta_{j}(w)=\delta_{Y}(w)/\delta_{X_{j}}(w).$
\end{proposition}

The idea is that any discontinuity in the outcome $\delta_Y(w)$ must be due only to the discontinuity in $X_j$, since all the other treatments are continuous across the threshold. Therefore, we can identify the conditional LATE of $X_j$, $\beta_j(w).$ 

\begin{figure}[H]
\caption{\textbf{First Stage, Unconditional}}%
\label{fig:unconditional}%
\vspace{-.5cm}
\begin{center}
\includegraphics[scale=0.58]{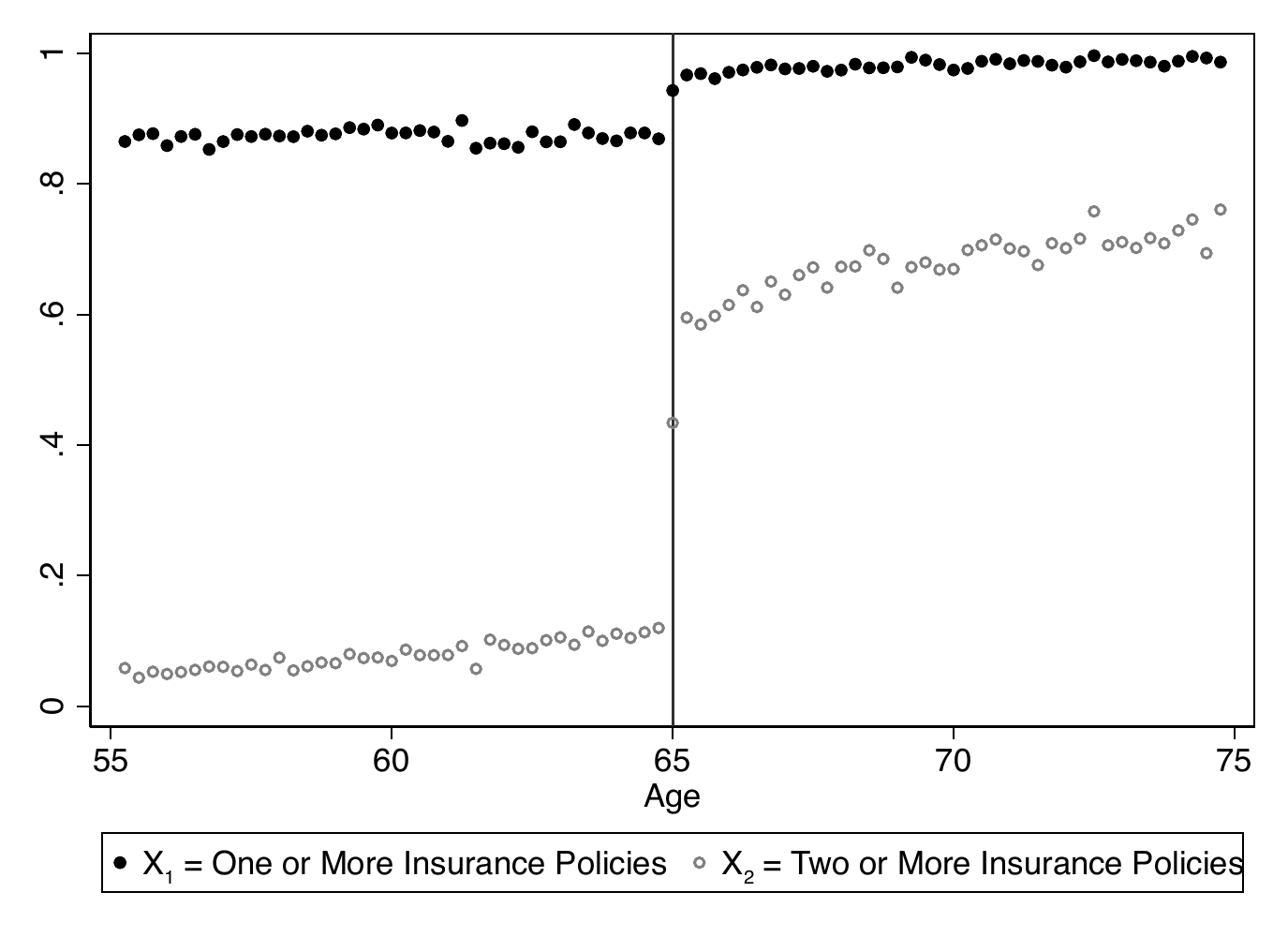}
\end{center}
\par
\vspace{-.5cm}
{\footnotesize \begin{singlespace} Note: The plot shows,  for each age (measured in quarters of a year), the percentage of people that have one or more insurances  (black dots) and two or more insurances (hollow dots).\end{singlespace}}
\end{figure}

Can Proposition \ref{Local} help us learn something about the treatment effects in our application? Figure \ref{fig:unconditional} shows that both $X_1$ (indicator for one or more insurance policies) and $X_2$ (indicator for two or more insurance policies) are  discontinuous across the threshold. Therefore, we cannot use Proposition \ref{Local} to learn about either of the LATEs $\beta_1$ or $\beta_2$. This is intuitive: the discontinuity in the outcome across the threshold is due to the discontinuities in both $X_1$ and $X_2$, and we cannot disentangle the two effects.

\begin{figure}[!htp]
\caption{\textbf{First Stage by Race}}%
\label{fig:race}%
\vspace{-.4cm}
\begin{subfigure}[b]{0.5\textwidth}
\caption{Whites}
\includegraphics[scale=0.58]{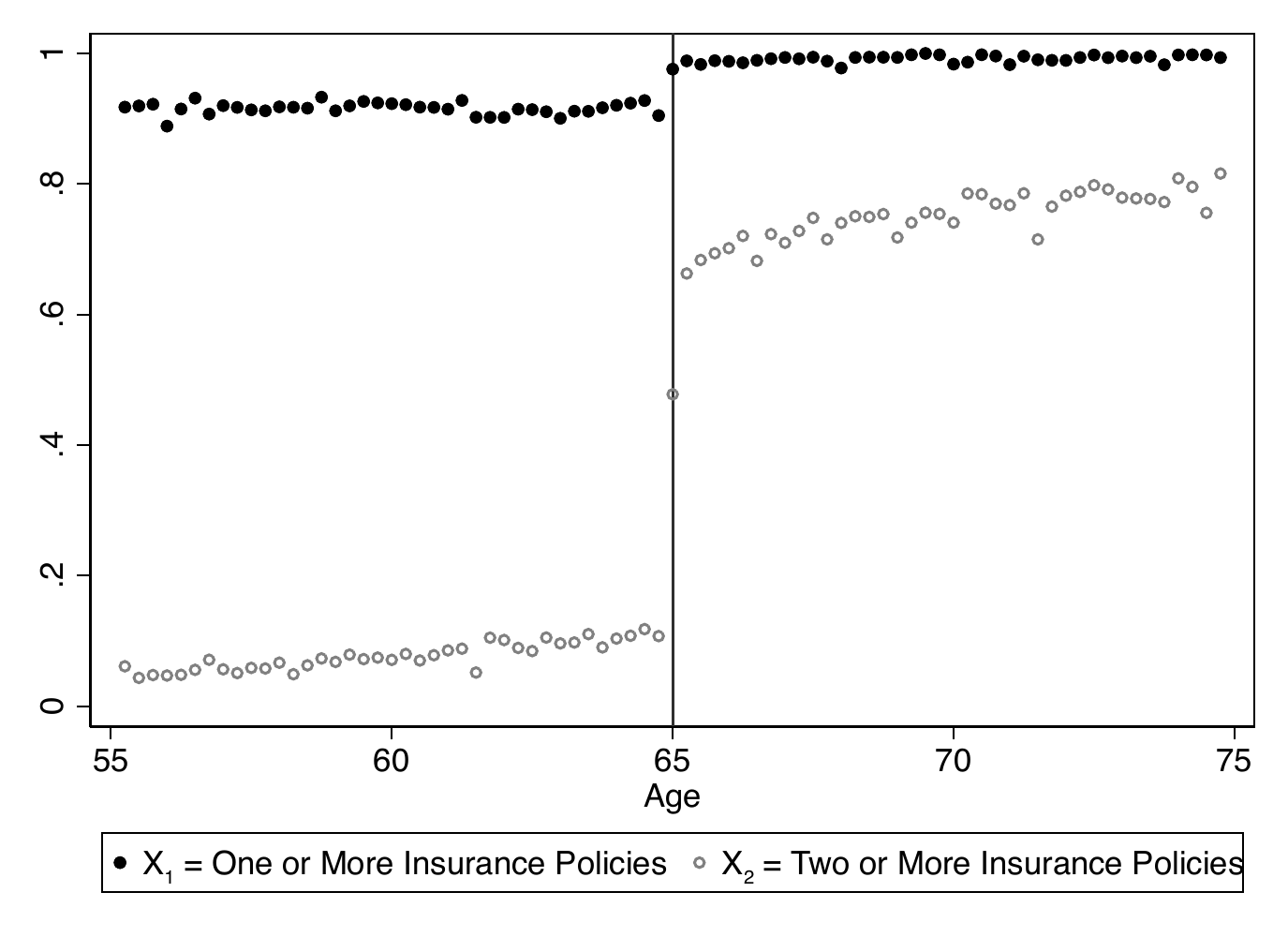}
\end{subfigure}
\begin{subfigure}[b]{0.5\textwidth}
\caption{Minorities}
\ \includegraphics[scale=0.58]{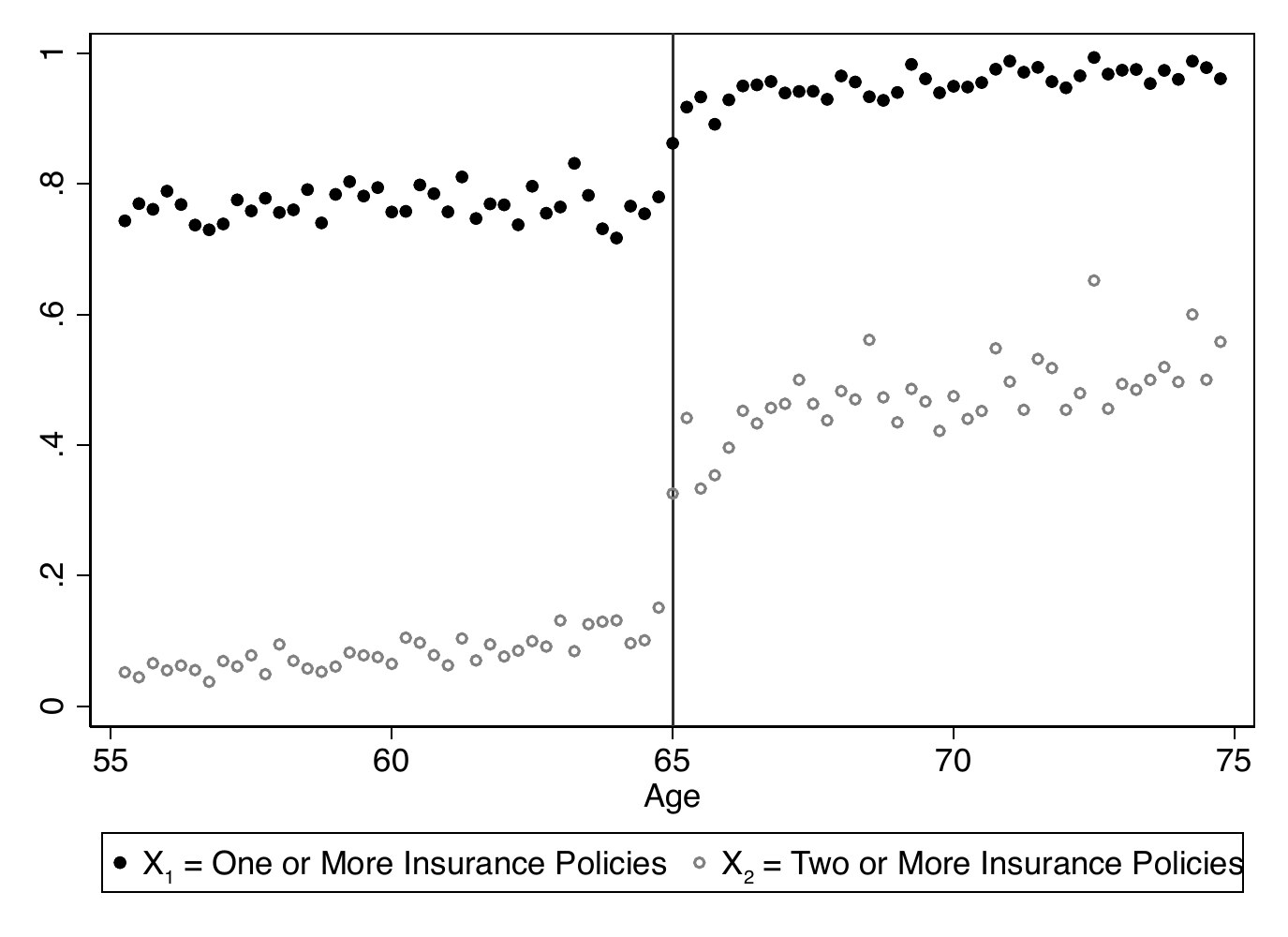}
\end{subfigure}
\par
{\footnotesize \begin{singlespace} Note: The plot shows,  for each age (measured in quarters of a year), the percentage of people with $W$ as described in the plot caption that have one or more insurances  (black dots) and two or more insurances (hollow dots).\end{singlespace}}
\end{figure}

Perhaps we can learn about some of the conditional LATEs. For example, let us consider the first stages conditional on $W=$ Race, which can be seen in Figure \ref{fig:race}. Unfortunately, for both whites (Panel (a)) and minorities (Panel (b)), both $X_1$ and $X_2$ are discontinuous, so we cannot learn anything about any of $\beta_1(\text{WH})$, $\beta_2(\text{WH})$, $\beta_1(\text{MIN})$,  or $\beta_2(\text{MIN})$. We have the same problem when $W=$ Education, which means that we cannot learn anything about $\beta_1(\text{COL}), \beta_2(\text{COL}), \beta_1(\text{HS}), \beta_2(\text{HS}), \beta_1(\text{DRP})$ or $\beta_2(\text{DRP})$ either.

\vspace{.2cm}

\begin{figure}[!htp]
\caption{\textbf{First Stage by Race and Education}}%
\label{FirstStageRaceEduc}%
\begin{subfigure}[b]{0.5\textwidth}
\caption{Whites with at least some college}
\includegraphics[scale=0.58]{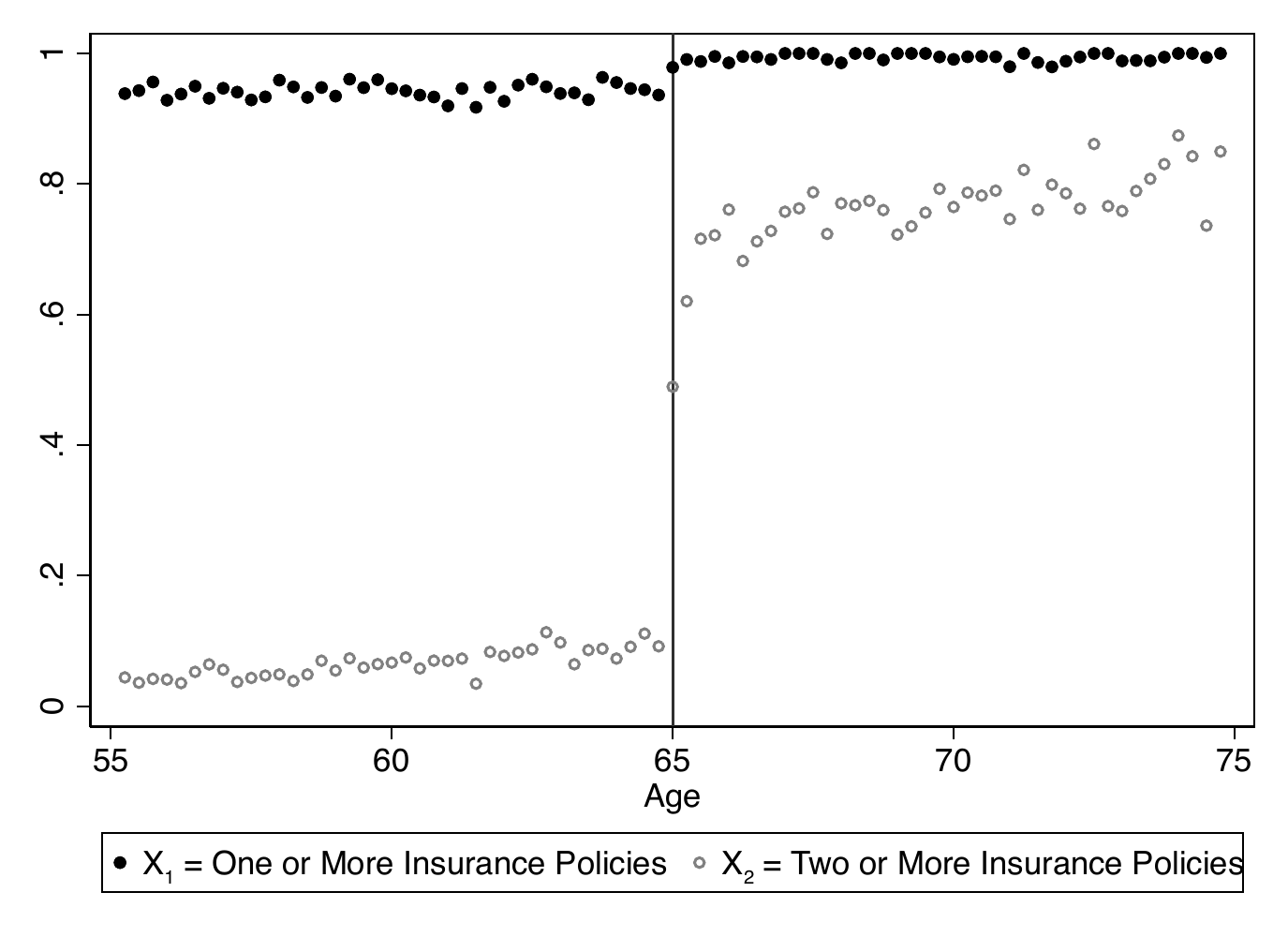}
\end{subfigure}
\begin{subfigure}[b]{0.5\textwidth}
\caption{Minorities with at least some college}
\ \includegraphics[scale=0.58]{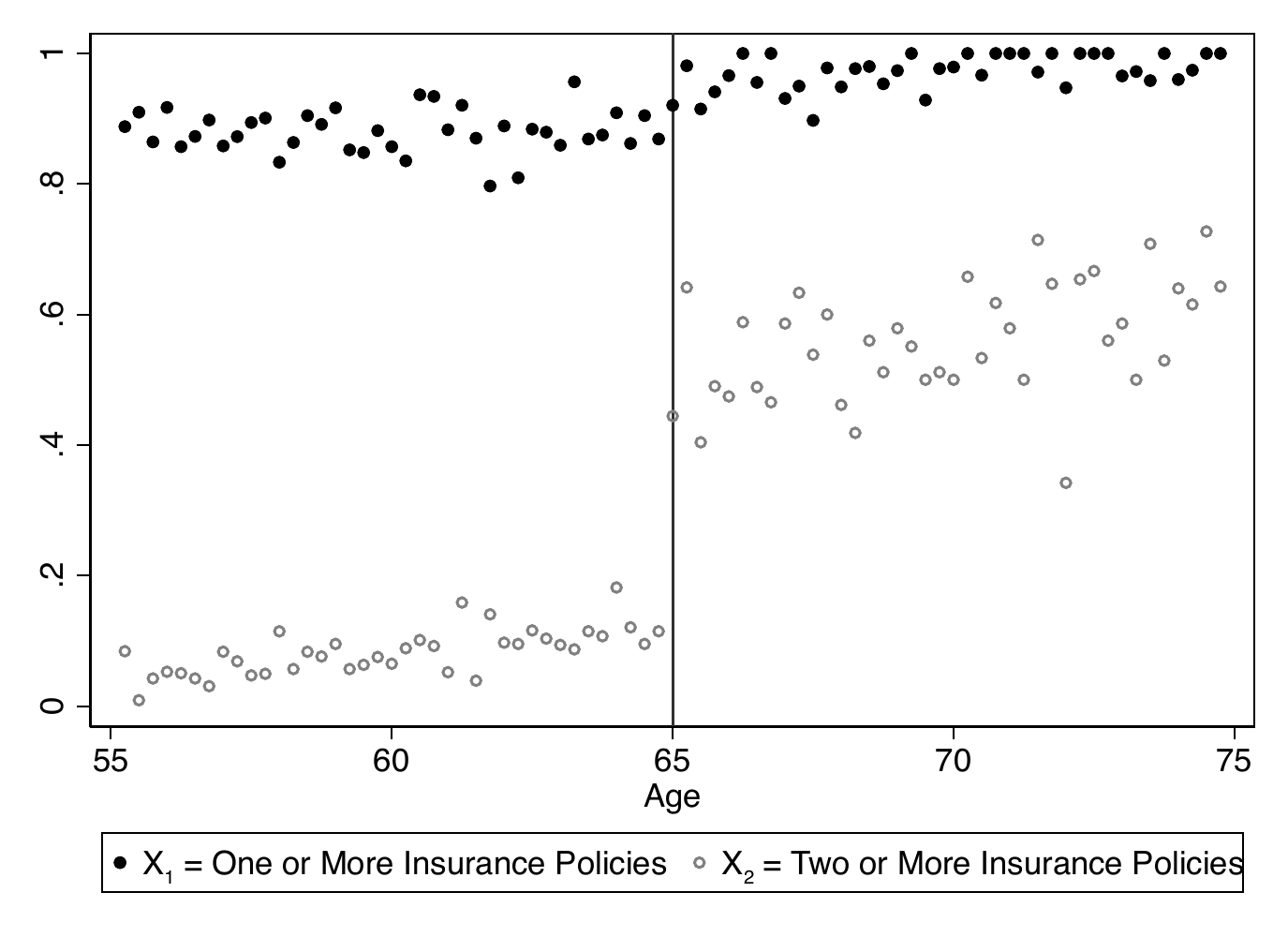}
\end{subfigure}
\begin{subfigure}[b]{0.5\textwidth}
\caption{Whites high school graduates}
\includegraphics[scale=0.58]{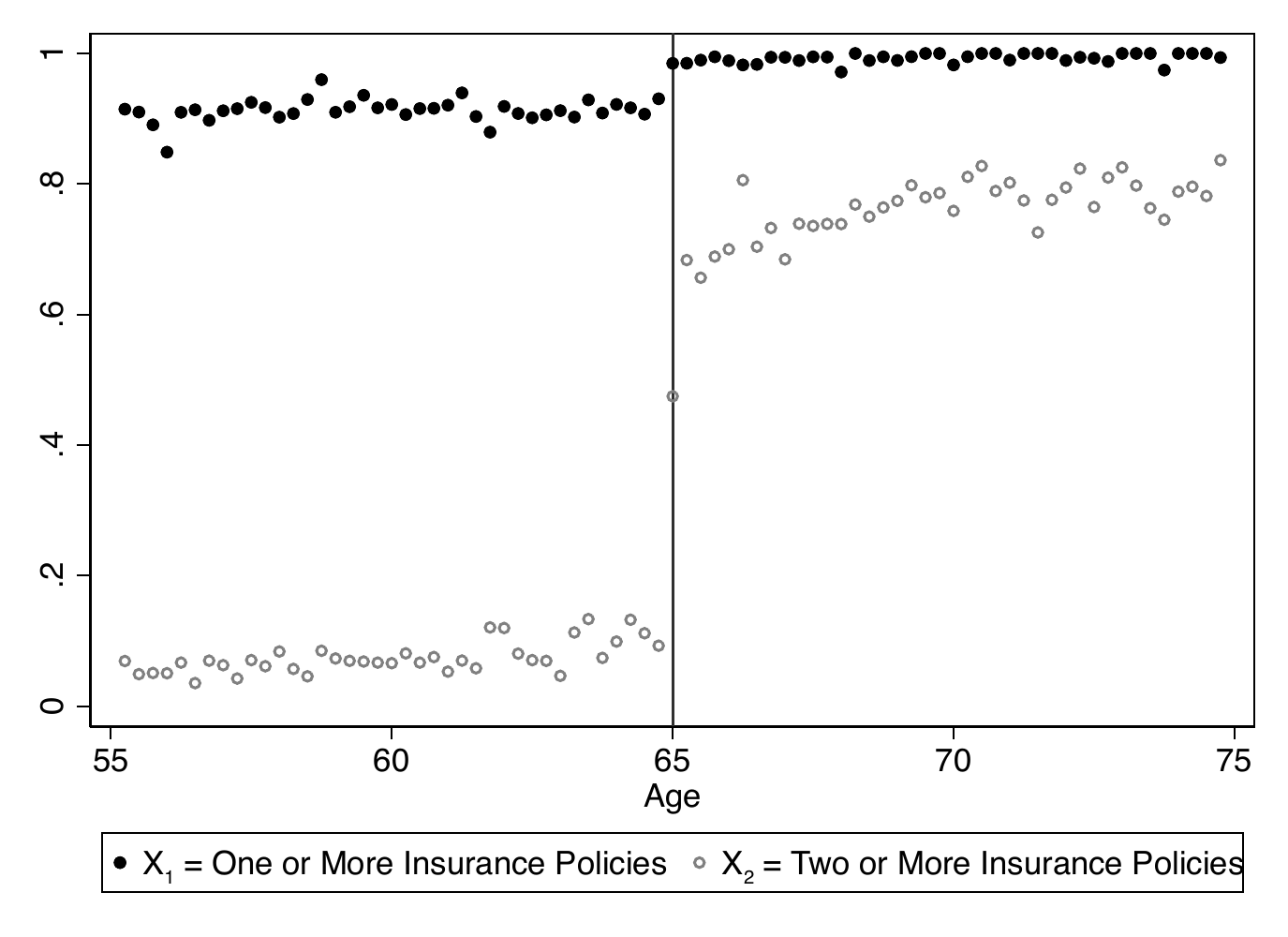}
\end{subfigure}
\begin{subfigure}[b]{0.5\textwidth}
\caption{Minorities high school graduates}
\ \includegraphics[scale=0.58]{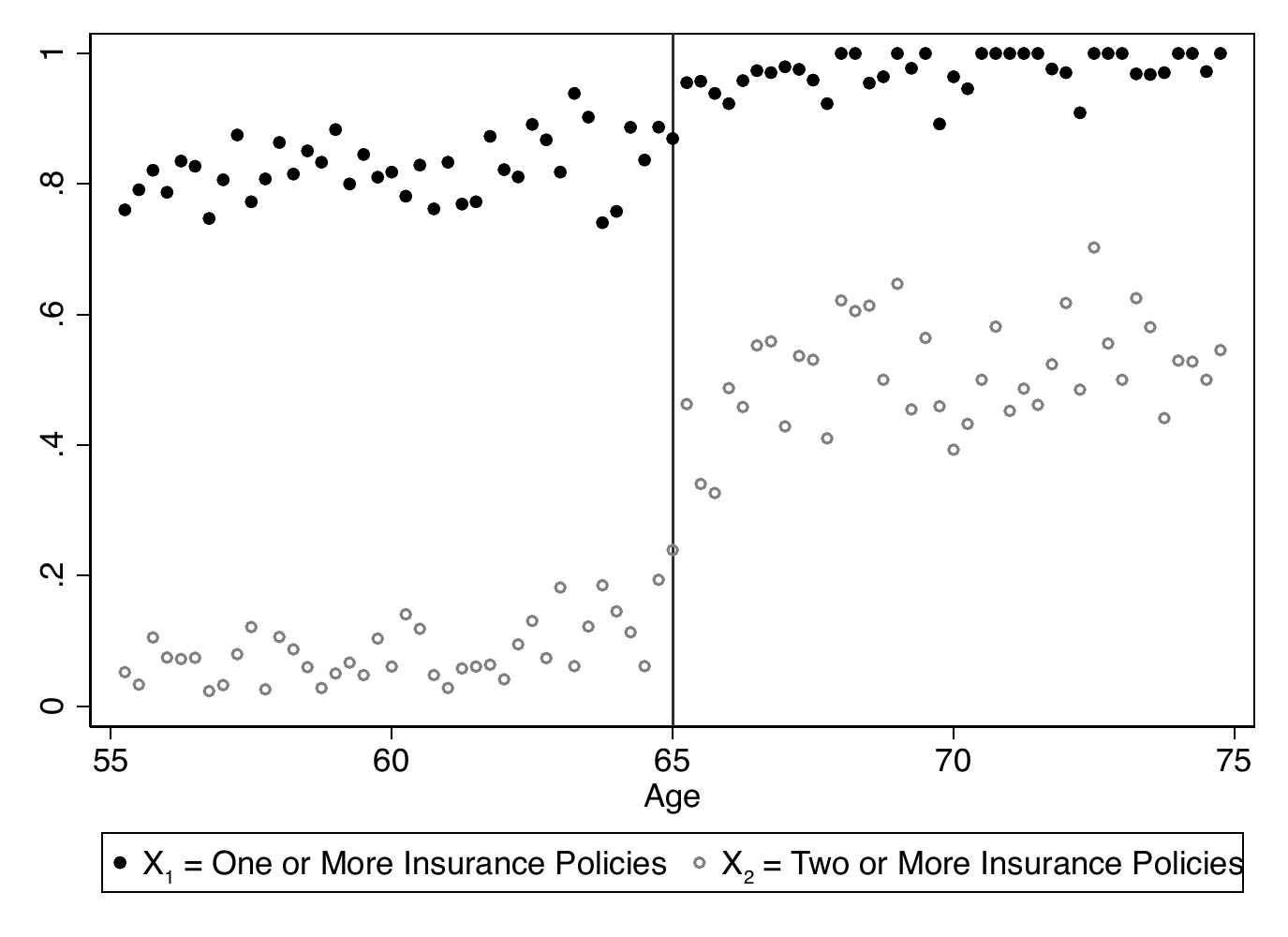}
\end{subfigure}
\begin{subfigure}[b]{0.5\textwidth}
\caption{Whites high school dropouts or less}
\includegraphics[scale=0.58]{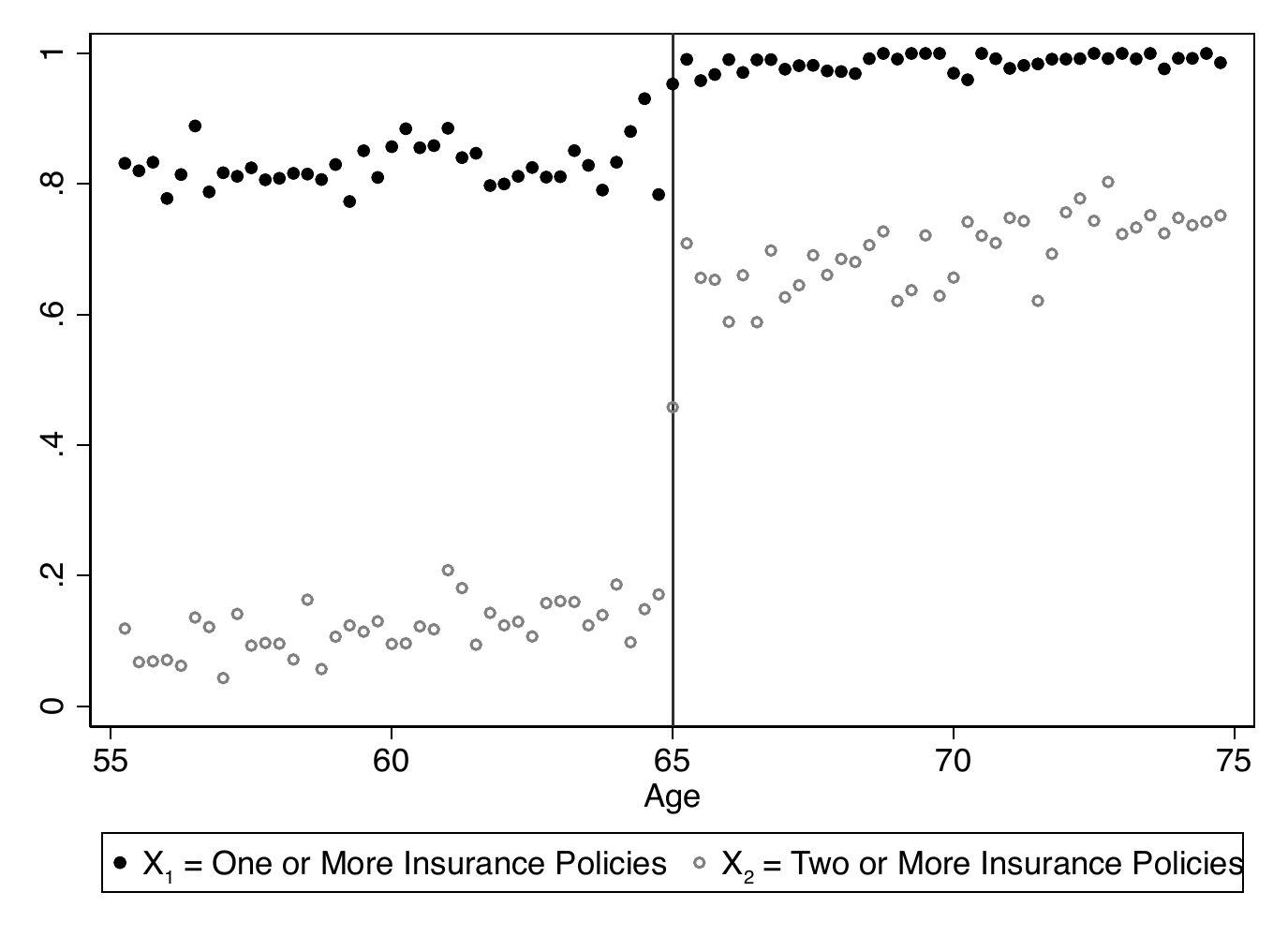}
\end{subfigure}
\begin{subfigure}[b]{0.5\textwidth}
\caption{Minorities high school dropouts or less}
\ \includegraphics[scale=0.58]{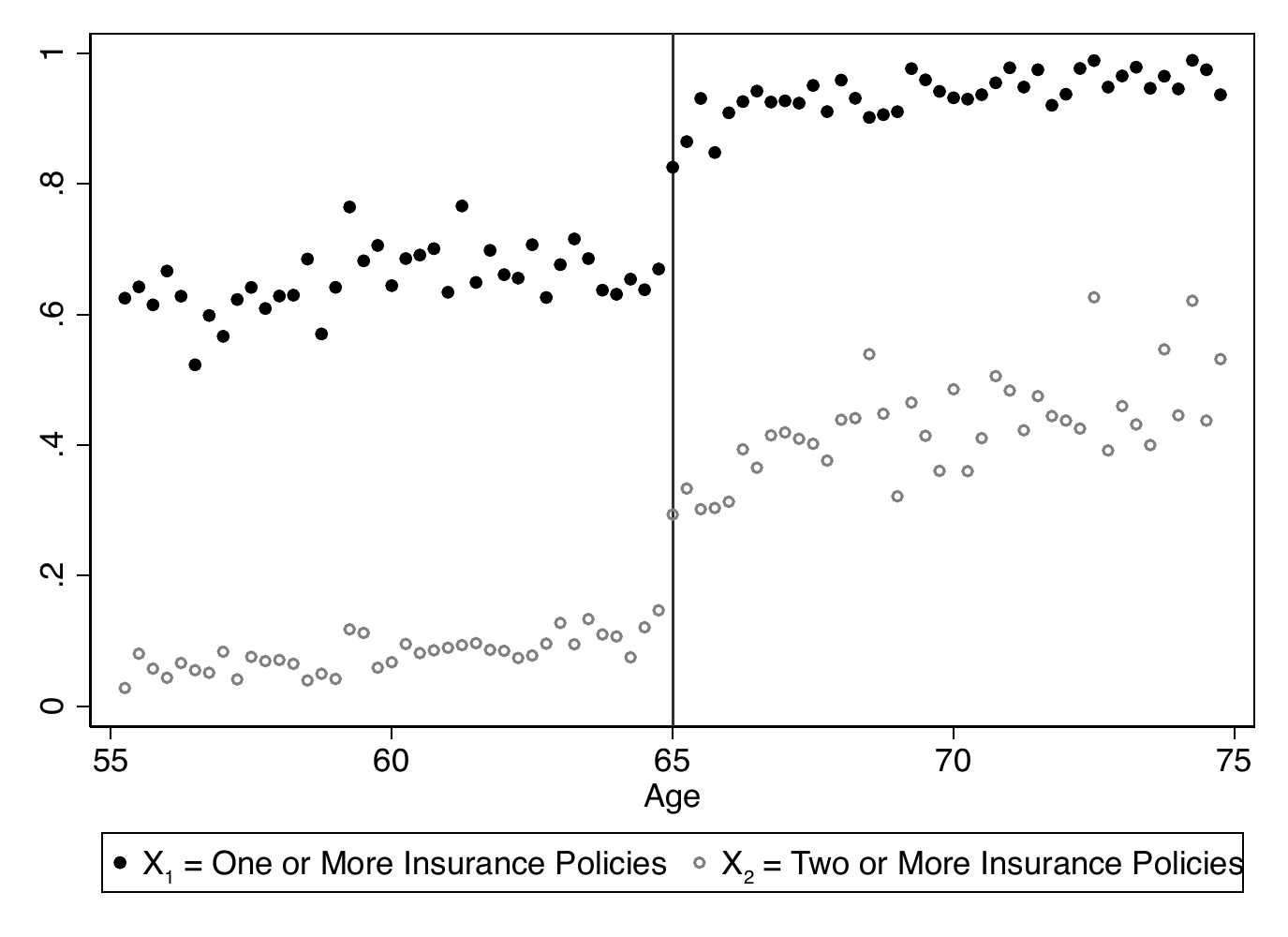}
\end{subfigure}
\par
{\footnotesize \begin{singlespace} Note: The plot shows,  for each age (measured in quarters of a
year), the percentage of people with $W$ as described in the plot caption that have one or more insurances  (black dots) and two or more insurances (hollow dots).\end{singlespace}}
\end{figure}

Moving one step further and making  $W=$ Race and Education (Figure \ref{FirstStageRaceEduc}), we find one group which appears to have no discontinuity in the variable $X_{1},$ and a sizeable discontinuity in the variable $X_{2}$: whites with at least some college (Figure \ref{FirstStageRaceEduc}(a)).
 If that is case, then any discontinuity in the outcome for this group should reflect exclusively the change in $X_2,$ and thus 
$\beta_{2}(\text{WH,COL})$ could be identified as $\delta_{Y}(\text{WH,COL})/\delta_{X_{2}}(\text{WH,COL}).$ 

In reality, although $\delta_{X_{1}}(\text{WH,COL})$  seems small in Figure \ref{FirstStageRaceEduc}(a), it is nearly 5\% (0.048) and highly significant ($t=$11.34). Therefore,  $\delta_{X_1}(\text{WH,COL})$ is not equal to zero and  we cannot really identify $\beta_{2}(\text{WH,COL})$ using Proposition \ref{Local}. The situation is even worse for other  combinations of race and education. For all the other cases, the discontinuity in the outcome reflects substantial  discontinuities in both $X_1$ and $X_2,$ and we cannot disentangle the effect of going from zero to one insurance from the effect of going from one to two or more insurances. 

In general, the restriction $\delta_{X_{-j}}(w)=0$ may work for some  $w$ on a specific application, but it is unlikely that a large set of  LATEs $\beta_j(w)$ will be identified in this way. In fact, if $\beta_j(w)$ is identified using Proposition \ref{Local}, then $\beta_{s}(w)$ is not identified for all $s\neq j$. Without further assumptions and using Proposition \ref{Local}, in the best case scenario only one of the marginal effects could be identified for each $w$.

Next, we consider the identification of Weighted Local Average Treatment Effects (WLATEs) $E[\omega_{j}(W)\beta_{j}(W)]$, for non-trivial identified weights $\omega_{j}$. Although this quantity can only inform us about a specific average of LATEs across different values of $W$, it at least includes only marginal effects of $X_j$, $\beta_j(W)$, avoiding contamination from the marginal effects of $X_s$, $\beta_s(W)$, for $s \neq j$. Unfortunately, the next result shows that the WLATEs can only be identified under the same circumstances in which Proposition \ref{Local} can also be used, and thus these types of averages are not any easier to identify than the LATEs themselves.

\begin{proposition}
\label{WATE}Under Assumption \ref{assumption structural RDD}, $E[\omega
_{j}(W)\beta_{j}(W)]$ is identified $\iff$ $\omega_{j}(W)\delta_{X_{-j}}(W)=0$ a.s.
\end{proposition}

This proposition implies that $\omega_j(w)$ can be different from zero only when $\delta_{X_{-j}}(w)=0$. This means that the identified  WLATE $E[\omega
_{j}(W)\beta_{j}(W)]$ must be a linear combination only of the $\beta_j(w)$ such  that $\delta_{X_{-j}}(w)=0$. However, when  $\delta_{X_{-j}}(w)=0$, Proposition \ref{Local} guarantees that $\beta_j(w)$ is identified. Therefore, Proposition \ref{WATE} does not identify any new object that we could not identify with Proposition \ref{Local} already. 

Proposition \ref{WATE} can thus be seen as an impossibility result: if one cannot apply Proposition \ref{Local} and thus identify $\beta_j(W)$ for some value of $W$ (which may be constant), then one cannot identify any WLATE  either. This result confirms the bleak state of affairs in the identification of LATEs in the multivalued setting. 

\subsection{Identification of Weighted Averages of All The Effects}\label{sec:twlate}\label{sec:wlate}
In this section we show that although it may not be possible to identify the LATEs $\beta_j(W)$, or even the WLATEs $E[\omega_{j}(W)\beta_{j}(W)]$ for useful weights,  it is possible to identify weighted averages $E[\omega(W)'\beta(W)]$, that combine the LATEs of all $X_j$, for non-trivial, identified multivariate weights $\omega(W)=(\omega_{1}(W),...,\omega_{d}(W))'$. We call these TWLATEs, where the $T$ stands for ``total".  The next proposition gives necessary and sufficient conditions for the identification of these objects.

\begin{proposition}
\label{MWATE}Under Assumption \ref{assumption structural RDD}, $E[\omega(W)'\beta(W)]$ is identified $\iff$ $\omega_{j}(W)=a(W)\delta_{X_{j}}(W)$ a.s. for some $a(W)$ and all $j=1,...,d$.
\end{proposition}

\noindent Proposition \ref{MWATE} implies that the only TWLATEs that can be identified are those with weights that are proportional to the first stages. Identification is constructive, in the sense that it is straightforward to build an estimator based on this identification result, namely, the sample analog of $E[a(W)\delta_{Y}(W)]$.

A key concern with TWLATEs is that they aggregate marginal treatment effects for different treatment values. For instance, if we want to identify marginal LATEs $\beta_j$, the TWLATEs will also include the marginal LATEs $\beta_s$ for $s\neq j$. Is it possible to identify TWLATEs that ``separate in expectation" the marginal LATEs? In other words,  although it is not possible to separate the LATEs $\beta_j$ from  $\beta_{s}$ for $s\neq j,$  can we weight the conditional LATEs  so that the identified TWLATE gives an expected weight of 1 to $\beta_{j}(W)$ and an expected weight of zero to all $\beta_{s}(W)$ for all  $s\neq j$? Moreover, can we  do this for all $j=1,...,d$?

Specifically, if we want to separate in expectation the effects of $X_j$, for each $j$, we need a matrix, $\omega$, of weight functions $\{\omega_{js}\}_{j,s=1}^d$ such that $E[\omega_{jj}(W)]=1$ and $E[\omega_{js}(W)]=0$, for $s\neq j$ (i.e. $E[\omega(W)]=I,$ where $I$ denotes the $d\times d$ identity matrix). For such  matrix of weights $w$, we  denote the corresponding vector of TWLATEs $\bar{\beta}^{\omega}_W=(\bar{\beta}^{\omega}_{1,W},\dots,\bar{\beta}^{\omega}_{d,W})'=E[\omega(W)\beta(W)]$, where each element separates the effects of one of the treatment values. 

To better understand the notation and the requirements of separation in expectation, consider our application. Let $W=$ Race, and $p(\text{WH})$ and $p(\text{MIN})$ be the marginal probabilities of $W=$ WH and $W=$ MIN, respectively. Then, the TWLATE that separates in expectation the effect of $X_1$  is the weighted mean 
\begin{align*}
    \bar{\beta}^{\omega}_{1,\text{Race}}&=p(\text{WH})\omega_{11}(\text{WH})\beta_{1}(\text{WH})+p(\text{MIN})\omega_{11}(\text{MIN})\beta_{1}(\text{MIN})
    \\
    &+p(\text{WH})\omega_{12}(\text{WH})\beta_{2}(\text{WH})+p(\text{MIN})\omega_{12}(\text{MIN})\beta_{2}(\text{MIN}),
\end{align*} 
where $p(\text{WH})\omega_{11}(\text{WH})+p(\text{MIN})\omega_{11}(\text{MIN})=1$ and $p(\text{WH})\omega_{12}(\text{WH})+p(\text{MIN})\omega_{12}(\text{MIN})=0$. 

Note that by Proposition \ref{MWATE}, to achieve identification we must have $\omega_{js}(W)=a_j(W)\delta_{X_s}(W)$, and by Assumption \ref{assumption structural RDD}(iii) the $\delta_{X_s}(W)$  must all be non-negative. Therefore, in order to achieve separation in expectation  ($E[a_{j}(W)\delta_{X_s}(W)]=0$), $a_j(W)$ must assume both positive and negative values. Thus, to summarize, in order to achieve separation of different marginal treatment effects, Propositions \ref{Lack}, \ref{Local} and \ref{WATE} imply that separation in expectation is generally the only option, and Proposition \ref{MWATE} implies that this is only possible if both positive and negative weights are used for weighting the LATEs of all treatment variables.

Proposition \ref{MWATE} and the requirement of separability in expectation imply that $a(W)$ must satisfy $E[a(W)\delta_{X}(W)']=I$. In general, there are infinite $a(W)$'s that  achieve this. These $a$'s can be constructed by finding a $d$-dimensional function $b(w)$ such that $E[b(W)\delta_{X}(W)']$ is non-singular, and then setting $a(w)=E[b(W)\delta_{X}(W)']^{-1}b(w)$. This construction is only possible if the following assumption holds (since otherwise   $E[b(W)\delta_X(W)']$ would be singular for all $b$).

\begin{assumption}
\label{Relevance} $E[\delta_{X}(W)\delta_{X}(W)^{\prime}]$ is positive
definite.
\end{assumption}
This assumption  requires that  the first stage vectors are linearly independent. This condition immediately rules out a constant $W,$ because in that case $E[\delta_{X}(W)\delta_{X}(W)^{\prime}]=\delta_{X}\delta_{X}^{\prime}$ (the product of the unconditional first stages) which is a $d\times d$ matrix resulting from the product of a $d\times 1$ vector with another $1\times d$ vector, and thus can have a rank of at most one.
In fact, if $W$ assumes a finite number of values, $q$, this assumption  rules out all $W$ which assume less than $d$ values, so $q\geq d.$\footnote{The result follows from the fact that the rank of two matrices is subadditive. If $W$ assumes $q$ values, then $E[\delta_{X}\delta_{X}^{\prime}]=\sum_{l=1}^{q} \delta_{X}(w_l)\delta_{X}(w_l)'P(W=w_l).$ For a given $w_l,$ the rank of $\delta_X(w_l)\delta_X(w_l)'$ is at most 1 (because $\text{rank}(AA')=\text{rank}(A).$) Therefore, if $W$ assumes $q$ values, the rank of the sum must be at most equal to $q,$ and if $q<d,$ the expectation will not be full-rank.\label{subad}} Therefore, this implies that the minimum number of conditional LATEs $\beta_s(W)$ inside of $\bar{\beta}^{\omega}_{j,W}$ is $d^2.$ In our application, this number is four.

Assumption \ref{Relevance} is testable. In fact, if the support of  $W$ is finite then Assumption \ref{Relevance}  holds if and only if there are $d$ values of $W$ for which the $d$ vectors  $\delta_X(W)$ are linearly independent.\footnote{This follows from the fact that the sum of a positive definite matrix and a positive semi-definite matrix is always positive definite (suppose that $A$ is positive definite and $B$ is positive semi-definite, then for all $v\neq 0,$ $v'(A+B)v\geq 0,$ and suppose that $v'(A+B)v=0,$ then $B$ would be negative definite, which is absurd.) Say that the  $\delta_X(w_l)$ are linearly independent for $l=1,\dots,d,$ then  $A=\sum_{l=1}^d \delta_X(w_l)\delta_X(w_l)'P(W=w_l)$ is positive definite. Since $B=\sum_{l=d+1}^{q} \delta_X(w_l)\delta_X(w_l)'P(W=w_l)$ is positive semi-definite, the result follows. \label{foot:pos}} In our application, this means that all that is required for Assumption \ref{Relevance} to hold is that two of the conditional first stage vectors are linearly independent. This is obviously true for $W=$ Race, as can be seen in Figure \ref{fig:race}, in which the sizes of the discontinuities of $X_1$ and $X_2$ across the threshold for whites and minorities are clearly not proportional. For $W=$ Race and Education, it is enough to compare the first stage discontinuities for whites with some college, Figure \ref{FirstStageRaceEduc}(a), with the first stage discontinuities for the minorities who are high school dropouts, Figure \ref{FirstStageRaceEduc}(f), which  are clearly not proportional.

The following result establishes that Assumption \ref{Relevance} is indeed necessary for separability in expectation. Moreover, Assumption \ref{Relevance} is also sufficient.

\begin{proposition}\label{cor:total}
Let Assumption \ref{assumption structural RDD} hold. Assumption \ref{Relevance} is necessary and sufficient for the identification of $E[\omega(W)\beta(W)]$ with weights $\omega$ such that  $E[\omega(W)]=I$. 
\end{proposition}

In order to see how  Assumption \ref{Relevance} is sufficient for separability in expectation, note that if $b(w)=\delta_X(w)$, then, $E[b(W)\delta_{X}(W)']$ is non-singular by Assumption \ref{Relevance}. Therefore, the weights
\begin{equation}\label{eq:weights}
\omega(w)=a(w)\delta_X(w)'=E[\delta_{X}(W)\delta_X(W)']^{-1}\delta_X(w)\delta_X(w)'
\end{equation} satisfy Proposition \ref{MWATE} and achieve separation in expectation.

Henceforth, we denote the specific vector of TWLATEs which uses the weights defined in equation  \eqref{eq:weights} as $\bar{\beta}_W=\left(E[\delta_{X}(W)\delta_{X}^{\prime}(W)]\right)  ^{-1}E[\delta_{X}(W)\delta_{X}^{\prime}(W) \beta(W)]$, with its $j$-th element denoted as $\bar{\beta}_{j,W}.$ Note that the weights are identified and can be estimated directly from the data by substituting the population by sample quantities. In Section \ref{RDDestimation} we discuss how to estimate these TWLATEs. The technique is rather simple and can be programmed into packaged software as a 2SLS regression.

To better understand the weights, note that we can write the corresponding TWLATEs as
\[
\bar{\beta}_W=\arg\min_{\gamma}E[\left(  \beta(W)-\gamma\right)  ^{\prime}\delta
_{X}(W)\delta_{X}^{\prime}(W)\left(  \beta(W)-\gamma\right)].
\]
The least squares objective function weights more heavily covariates for which the
first stages are larger. This means that, although the weights of $\bar{\beta}_{j,W}$ are calibrated so as to ``separate in expectation'' the  effects of $X_j$ from the effects of all the  other variables, the relative magnitudes of the weights are also  influenced by the size of the first stages. 

For example, in our application, if $W=$ Race, the weights with the highest expected magnitude in $\bar{\beta}_{1,W}$ are the ones multiplying the LATEs conditional on $W=\text{MIN}$, and indeed in Figure \ref{fig:race} we can see that the highest discontinuity for $X_1$ is for minorities (Panel (b)). Conversely, the weights with the highest expected magnitude in $\bar{\beta}_{2,W}$ are the ones multiplying the LATEs conditional on  $W=\text{WH}$, and indeed in Figure \ref{fig:race} we can see that the highest discontinuity for $X_2$ is for whites (Panel (a)).

If $W=$ Race and Education, the weights  with the highest expected magnitudes in $\bar{\beta}_{1,W}$ are the ones multiplying the LATEs conditional on $W=(\text{MIN,DRP}).$ Indeed, among all values of $W$ in Figure \ref{FirstStageRaceEduc}, the highest discontinuity in $X_1$ is for that group, which can be seen in Figure \ref{FirstStageRaceEduc} (f). Among all treatment effects included in $\bar{\beta}_{2,W},$ the weights with the highest expected  magnitude are the ones multiplying the LATEs conditional on $W=(\text{COL,WH}).$ As can be seen in Figure \ref{FirstStageRaceEduc} (a), the highest discontinuity in $X_2$ is for that group.

Note that we may also be able to partition the set of covariates and identify conditional TWLATEs $\bar{\beta}_{W,R} (r) = E[\omega(W,R)\beta(W,R) |R=r]$. The advantage of such conditional quantities is that they restrict the average only to the LATEs of that subgroup, with no contamination from the LATEs of other subgroups. Provided Assumptions \ref{assumption structural RDD} and \ref{Relevance} hold conditional on $R=r$, we can apply previous results conditional on $R=r$ to identify $\bar{\beta}_{W,R} (r).$

For example, in our application,  let $W=$ Race, and  $R=$ Education. Assumption \ref{Relevance} holds conditional on $R=$ COL: Figures \ref{FirstStageRaceEduc} (a) and (b) show that $\delta_{X}(\text{WH,COL})$ and $\delta_{X}(\text{MIN,COL})$ are clearly linearly independent. Let $p=P(W=\text{WH}|R=\text{COL}),$ then
\begin{align*}
    \bar{\beta}_{1,\text{Race,Educ.}}(\text{COL}) &=p\cdot\omega_{11}(\text{WH,COL}) \beta_1(\text{WH,COL})+(1-p)\cdot \omega_{11}(\text{MIN,COL})\beta_1(\text{MIN,COL})
    \\
    &+p\cdot \omega_{12}(\text{WH,COL})\beta_2(\text{WH,COL})+(1-p)\cdot \omega_{12}(\text{MIN,COL})\beta_2(\text{MIN,COL}).
\end{align*}
The weights are such that $p\cdot\omega_{11}(\text{WH,COL})+(1-p)\cdot \omega_{11}(\text{MIN,COL})=1$ and $p\cdot\omega_{12}(\text{WH,COL})+(1-p)\cdot \omega_{12}(\text{MIN,COL})=0.$ Moreover, all the treatment effects in $\bar{\beta}_{1,\text{Race,Educ.}}(\text{COL})$ refer to the people with at least some college, with no contamination at all from the treatment effects of those with high school degree or less. Analogously, we can identify $\bar{\beta}_{1,\text{Race,Educ.}}(\text{HS})$ and $\bar{\beta}_{1,\text{Race,Educ.}}(\text{DRP})$, since in both cases Assumption \ref{Relevance} conditional on these two subgroups is clearly valid (see Figures \ref{FirstStageRaceEduc} (c) and (d) for $R=$ HS, and Figures \ref{FirstStageRaceEduc} (e) and (f) for $R=$ DRP).

\section{Identifying the Marginal Local Average Treatment Effects} \label{sec:exclusion}

It should be clear from the previous section how difficult it is to identify useful functions of the LATEs with parsimonious assumptions. The source of the problem is the dimension of the space of treatment effects, which is just too large. We have $d\cdot q$ treatment effects and only $q$ identification equations. In order to identify the specific treatment effects it is  necessary to reduce the space of treatment effects to at most $q.$  

In this section we propose a simple way to reduce the space of treatment effects: homogeneity of the LATEs on $W.$ We study how this assumption might be tested, and offer two strategies for relaxing it: conditional homogeneity  (Section \ref{sec:exclusionR}), and a flexible parametric approach (Section \ref{sec:parametric}). We also discuss how all these models may be tested as well as the trade-offs involved in choosing which strategy to employ.

\begin{assumption}\label{H} $\beta(W)=\beta$ a.s.
\end{assumption}

Assumption \ref{H}  says that conditioning the treatment effects on $W$ is the same as not conditioning on them, i.e. $\beta_{j}(W)=\beta_j$. This is in essence a separability condition which states that although $W$ is allowed to be a confounder (i.e. $W$ may be correlated with the intercept of the treatment equation \eqref{RC}), it is not allowed to be correlated with the treatment effects near the threshold.

To understand what Assumption \ref{H} entails, let us consider our application. Suppose that  we choose $W=$ Race. Then, Assumption \ref{H} requires that $\beta_j(\text{WH})=\beta_j(\text{MIN})=\beta_j$ for $j=1,2$. Specifically, it means that, on average, whites and minorities who were not insured but at 65 get covered by Medicare will change their behavior in the same way with respect to, for example,  delaying or rationing care for cost reasons, or deciding to go to a hospital. The same will be true also for whites and minorities  who had insurance before but at 65 get double coverage by Medicare and another insurance.  


Assumption \ref{H} automatically reduces the number of parameters to identify to $d$. Therefore,  as long as Assumption \ref{Relevance} holds, we can identify the LATEs $\beta_j.$ 

\begin{proposition}
\label{Identificationext} Let Assumptions
\ref{assumption structural RDD} and \ref{H} hold. Then, $\beta$ is
identified if and only if Assumption \ref{Relevance} holds.
\end{proposition}

Proposition \ref{Identificationext} actually states that Assumption \ref{Relevance} is the weakest possible relevance condition for the identification of the LATEs. In fact, under Assumptions \ref{assumption structural RDD}, \ref{Relevance} and \ref{H}, the TWLATEs $\bar{\beta}_W$ that separate in expectation (Section \ref{sec:wlate}) are equal to  the LATEs $\beta$, 
and we can write
\begin{equation}\label{identeq}
\beta=\left(E[\delta_{X}(W)\delta_{X}^{\prime}(W)]\right)  ^{-1}%
E[\delta_{X}(W)\delta_{Y}(W)].
\end{equation}

Note that this identification strategy  may lead to over-identification. We have $q$ identifying equations \eqref{rddlinear}, with $d$ unknown LATEs $\beta_j$, so we may have up to $q-d$ over-identifying restrictions for the identification of $\beta.$

In our application, we need to identify two LATEs, $\beta_1$ and $\beta_2$. If $W=$ Race, then  $q=2$ and thus there is no over-identification. However, if   $W=$ Race and Education, then $q=6$, and thus there are four over-identifying restrictions for the estimation of $\beta_1$ and $\beta_2$. If $W=$ Race, Education and Gender, then $q=12$ and thus there are 10 over-identifying restrictions. 

The trade-off is clear. When we add one element to $W,$ the assumption of homogeneous treatment effects on $W$ becomes stronger, but at the same time, the model becomes more over-identified. If $q$ is the number of values of the previous elements and the new element assumes $q'$ values, then there will be $q(q'-1)$ additional degrees of over-identification. This is a strength of this method: when we add more terms to $W$ the identifying assumption is more likely to be false, but at the same time our ability to test that assumption increases.

To understand the over-identification, suppose that Assumption \ref{H} is wrong. In this case, 
\begin{equation*}
    \delta_Y(W)=\beta'\delta_X(W)+(\beta(W)-\beta)'\delta_X(W),
\end{equation*}
and thus the moment conditions  $E[(\delta_Y(W)-\beta'\delta_X(W))W]=0$ are not satisfied. This is exactly what over-identification tests such as Sargan-Hansen's $J$-test are built to detect. As we will show in Section \ref{RDDestimation}, our estimators can be transformed into standard 2SLS estimators and thus performing the test in packaged software is as simple as running the regression under the assumed model and then selecting the option to display the over-identification test results.

In the context of an instrumental variable regression, a  rejection of the null in the $J$-test can be caused both by a problem with the validity of the instruments or the mispecification of the functional form. However, note that in the case of the RDD, the specification is not an issue, since we are estimating the limits nonparametrically, and the fundamental validity of the approach (Assumption \ref{assumption structural RDD}) is usually quite convincing. This means that a rejection can be more safely interpreted as a failure of the exclusion restriction $\beta(W)=\beta$ which we are aiming to test.\footnote{\label{foot:testing}
An alternative to the use of an over-identification test is to  perform direct tests in which we use the over-identification to estimate a  more general model, then test if Assumption \ref{H} holds.  The simplest strategy is to estimate a parametric random coefficients model as in Section \ref{sec:parametric}, and then test whether the coefficients of all terms but the constant are zero with an $F$-test. Another approach is to subdivide the vector $W$ in two parts, so $W=(W_1',W_2')',$ and estimate the nonparametric model from Section \ref{sec:exclusionR} treating $W_1$ as $W$ and $W_2$ as $R$. Then, test whether the $\beta(R)=\beta$ with an $F$-test. 

Although these tests have the right size under the null hypothesis, they may have low power. The general models are less over-identified, and thus are estimated with much less precision than the unconditional LATEs. Therefore, a lot of power may be lost simply because the test statistic is just too noisy in some applications. Moreover, if the more general model does not hold, which may well be the case, it is unclear what the estimator is identifying under the alternative hypothesis, and therefore the power may not increase with the heterogeneity of the LATEs.}

\subsection{Relaxing Assumption \ref{H}: a nonparametric model with an exclusion restriction}\label{sec:exclusionR}

Assumption \ref{H} may be relaxed  by conditioning on another set of covariates in the data. We propose the alternative assumption
\begin{assumption}\label{Hc}
$\beta(W,R)=\beta(R).    $
\end{assumption}

It is generally easier to argue that $W$ can be excluded conditionally than unconditionally. 
For example, in our application, suppose that $R=$ Education. Then, Assumption \ref{Hc} requires that $\beta_j(\text{WH},R)=\beta_j(\text{MIN},R)=\beta_j(R)$ for $R=$ COL, HS and DRP, and $j=1,2.$ Specifically, it means that, on average, whites and minorities who have the same educational level and were not insured but at 65 get covered by Medicare will change their behavior in the same way with respect to, for example,  delaying or rationing care for cost reasons, or deciding to go to a hospital. The same will be true also for whites and minorities  who have the same educational level and had insurance before but at 65 get double coverage by Medicare and another insurance. However believable this assumption is, it is certainly more likely that people with the same education and different races will take similar actions, than to believe that people with different races will take the same actions when we restrict nothing else, as in Assumption \ref{H}.

If Assumption \ref{Hc} holds, and Assumptions \ref{assumption structural RDD} and \ref{Relevance} hold conditional on $R$ a.s., then Proposition \ref{Identificationext} can be applied conditional on $R$, and
\begin{equation*}
\beta(R)=\left(E[\delta_{X}(W)\delta_{X}^{\prime}(W)|R]\right)  ^{-1}%
E[\delta_{X}(W)\delta_{Y}(W)|R] \quad \text{a.s.}
\end{equation*}
Then, the unconditional LATEs $\beta_j$ can be  identified per equation \eqref{eq:media}.

If $R$ assumes $q_R$ values, then there are now $d\cdot q_R$ LATEs to identify, and $q\cdot q_R$ restrictions, for a total of $(q-d)\cdot q_R$ over-identifying restrictions. Specifically, we have $q-d$ over-identifying restrictions in the identification $\beta(R)$, for each of the values $R$ assumes.

The estimator of $\beta(R)$ can be programmed as a 2SLS, as we show in the Appendix Section \ref{sec:heterogeneityW1}, and thus over-identification  and other tests can be performed analogously to the unconditional case.

\subsection{Relaxing Assumption \ref{H}: a parametric model}\label{sec:parametric}

This strategy transforms our model in equation \eqref{RC} into a parametric random coefficients model.
\begin{assumption}\label{T}
$\beta(W)=g(W,\theta)$ a.s.,
\end{assumption}
where $g$ is a function which is known up to a  finite parameter vector $\theta$ with  $p$ elements. There are $q$ identifying equations and $p$ parameters, and thus the restriction is that $p\leq q.$

 A simple example is the linear specification
\begin{equation}\label{linear}
   \beta(W)=\theta_0+\theta_1' \tilde{W},
   \end{equation}
where $\tilde{W}$ is a vector with $c$ elements which are functions of the elements of $W,$ and may include the elements of $W$ themselves, indicators of sets of values of $W,$ interactions, higher order terms, etc., and may even not include some elements of $W$ at all. $\theta_0$ is a $d\times 1$ vector and $\theta_1$ is a $d\times c$ matrix. Here $\theta=(\theta_0',\text{vec}(\theta_1)')'$ ($\text{vec}(A)$ transforms the columns of $A$ into a single column vector) and $p=d(1+c).$ The restriction is therefore $c\leq q/d-1.$ 
In our application, if $W=$ Race and Education,  $c\leq 6/2-1=2,$ which means that $\tilde{W}$ can have at most two elements.

The identification equations can be written as
\begin{equation*}
    \delta_Y(W)=\theta'[(1,\tilde{W})\otimes I]\delta_X(W).
\end{equation*}
Define $\tilde{X}=(X',\tilde{W}_1X',\dots,\tilde{W}_{c}X')',$ where $\tilde{W}_1,\dots,\tilde{W}_{c}$ are the elements of $\tilde{W}.$ Then $\delta_Y(W)=\theta'\delta_{\tilde{X}}(W)$, which is equivalent to a model with treatment variable $\tilde{X}$ which satisfies Assumption \ref{H}. 
All the assumptions and identification results in this paper also hold when the treatment variables are not binary. Therefore,  Assumption \ref{Relevance} can be tested, and the vector of parameters $\theta$ can be identified identically to equation \eqref{identeq}, but using  $\tilde{X}$ instead of $X.$ 
The conditional LATEs $\beta_j(W)$ are therefore identified by equation \eqref{linear}, and the unconditional LATEs $\beta_j$ are identified by equation \eqref{eq:media}.

 For a given $W,$ Assumption \ref{T} is less restrictive the more flexible is the specification of $\beta(W).$ The constraint is that $p\leq q,$ and it is tempting to  include as many variables and interactions as possible. At the extreme, we would specify $\tilde{W}$ as a vector of $q/d-1$ variables. However, if the specification is kept parsimonious and thus $p$ is small, there are $q-p$  over-identifying restrictions (under equation \eqref{linear} there are $q-d(1+c)$ over-identifying restrictions).
Therefore, our recommendation is that the specification be chosen less as a ``kitchen sink" and more as a well thought out model which is based as much as possible on institutional knowledge and economic theory. We can then use the over-identification generated by this approach to test the chosen model. 

Similarly to the nonparametric case, a sensible way to proceed is to estimate the treatment effects under the assumed model and then perform an over-identification test such as Sargan-Hansen's $J$-test.
An alternative to the use of an over-identification test  is to perform a  specification test, such as for example the modified Ramsey's RESET test for instrumental variable regression (\cite{PaganHall,PesaranTaylor}), which is also available for packaged software.

\subsection{How to choose the model}
If the objective is the identification of the unconditional LATEs, $\beta_j$, we believe that the more restrictive Assumption \ref{H} may often be the best choice. If the over-identification test does not reject for increasingly larger $W,$ while the estimates remain stable, it is sensible to stop seeking any more flexibility. Our Application in Section \ref{application} shows an example in which the results remain stable even after the p-values of the over-identification test start to get small.

If there is a wish to identify specific conditional LATEs, $\beta_j(W),$ or the confidence in Assumption \ref{H} for even a very conservative choice of $W$ is low, it is worth it to give up some over-identification for a more flexible model. In this case, how to choose between the nonparametric model with exclusion restrictions from Section \ref{sec:exclusionR} and the parametric random coefficients model from Section \ref{sec:parametric}?

These models have different strengths. To see this, consider how to model  $\beta(W,R)$.  
Under the nonparametric model, there is complete flexibility of the treatment effects with respect to $R$, but $W$ must be excluded. Conversely, under the parametric model  $\beta(W,R)$ does not need to exclude $W,$ at the cost of a less flexible function in $R.$

For instance, in our application, suppose that $W=$ Race and $R=$ Education. In the nonparametric approach, we assume $\beta(\text{Race, Education})=\beta(\text{Education})$.  In the parametric approach with a linear specification, the number of terms in $\tilde{W}$ is $c\leq  2.$ We can therefore include  Race in the specification, for example $\beta(\text{Race,Education})=\gamma_0+\gamma_{11}1(\text{Race}=\text{WH})+\gamma_{12}1(\text{Education}\geq \text{HS}).$ Neither assumption is weaker than the other in principle.

\section{Estimation}\label{RDDestimation}\label{sec:homogeneityW1}
We propose an estimator that delivers the TWLATEs developed in Section \ref{sec:wlate} or that, under the models of Section \ref{sec:exclusion}, delivers the 
LATEs. The estimator retains the structure  as similar as possible to the standard RDD while allowing for the inclusion of covariates and multiple treatments. This estimator can be programmed as a Two-Stage Least Squares (2SLS) regression.

For simplicity, we present here the unconditional case. This case covers the unconditional TWLATEs  from Section \ref{sec:wlate}, the unconditional LATEs  from Assumption \ref{H} in Section \ref{sec:exclusion}, and the  coefficients  of any linear model using  Assumption \ref{T} in Section \ref{sec:parametric}.  The conditional case can be seen in Appendix \ref{sec:heterogeneityW1}. It covers the conditional TWLATEs  from Section \ref{sec:wlate}, and the conditional LATEs  from Assumption \ref{Hc} in Section \ref{sec:exclusionR}. Consider the following definitions.
\begin{enumerate}
    \item Recenter $Z$ so that  $z_0=0$ (subtract $z_0$ from all $Z_i$ so that the threshold now falls at zero). Define $D=1(Z\geq 0).$
\item Define $C=(1,{W}',Z,D\cdot Z,Z\cdot {W}',D\cdot Z\cdot {W}')^{\prime}.$

\item Let $k_{h_{n}}(Z)=k(Z/h_{n}),$ where $k$ is a kernel function, and $h_{n}$ is a
bandwidth parameter satisfying some standard conditions in Section
\ref{AppendixSectionRRD} in the Appendix.

\end{enumerate}
We formalize our approach as a weighted 2SLS regression with first-stage regressions
\begin{equation*}
    {X}_{ij}=\pi_{0j}D_i+\pi_{1j}'{W}_i\cdot D_i+\alpha_{j}'C_i+\varepsilon_{ij}^x, \quad j=1,\dots,d
\end{equation*}
second stage regression
\begin{equation*}
    \hspace{-2.5cm}{Y}_{i}=\beta'\hat{{X}}_i+\eta'C_i+\varepsilon_{i}^y,
\end{equation*}
and weights equal to $k_{h_{n}}(Z_i).$ In other words, we propose running a weighted  2SLS regression of $Y$ on ${X}$ using $D$ and $D\cdot {W}$ as IVs,  $C$ as exogenous controls, and the $k_{h_n}$ as weights.\footnote{In Stata, the code is ``ivregress 2sls Y C (X = D D${\text{W}}$) [k]".}  Standard errors and tests obtained  directly from packaged software can be used for inference.

In general, the estimated coefficient of $X_j,$ $\hat{\beta}_j,$ estimates the   TWLATE $\bar{\beta}_{j,W}.$ Under the assumption $\beta(W)=\beta$ (Assumption \ref{H} in Section \ref{sec:exclusion}),  $\hat{\beta}_j$ estimates the LATE ${\beta}_j.$ 
Under the parametric random coefficients model $\beta(W)=\theta_0+\theta_1\tilde{W}$ (Section \ref{sec:parametric}),  estimation should be performed identically, substituting $X$ by $\tilde{X}$ and $\beta$ by $\theta$. In this case, $\hat{\theta}_j$ estimates the $j$-th element of the vector $\theta.$ 

Although the estimator is implemented as a weighted 2SLS regression, it is actually a local linear regression  on the running variable $Z$ on each side of the threshold on both the first stage and second stage regressions.  To see this, note that, similarly to the standard RDD, this estimator weighs observations by the  kernels, which restricts the sample only to the proximity of the threshold, and includes in the regression (inside of $C$) the terms $1,$ $Z$ and $D\cdot Z.$ This allows the expected outcome to vary with the running variable, and also to change derivative across the threshold. However, different from the standard RDD, this estimator includes ${W},$ $Z\cdot {W}$ and $D\cdot Z\cdot {W}$ in $C,$ which allows the outcome to vary with the running variable differently for every element of ${W},$ and also to change slope across the threshold differently for every element of ${W}.$

We use the local linear estimator for its excellent boundary properties (\cite{Imbens_Lemieux}), as does the current RDD literature. Because we use the local linear method on a single running variable $Z,$ and a single threshold, some of what is already known in the RDD literature can be applied directly here. For instance, we know that triangular kernels tend to perform better in boundary estimation (\cite{Lee_Lemieux}), and thus should be chosen here as well. We know that the use of controls can improve the efficiency of the estimators (\cite{Calonico_Cattaneo_Farrell_Titiunik}), and thus, if there are leftover covariates in the data which are not used in $W,$ they could be included as additional terms in the vector $C.$ Finally, methods available for optimal bandwidth choice in the standard RDD (e.g. \cite{Imbens_Kalyanaraman, Calonico_Cattaneo_Titiunik}) could be  adapted to this estimator, although the specifics are left for future research.

The following result establishes the asymptotic normality of the proposed estimator. Its  regularity conditions and proof are given in the Appendix
Section \ref{AppendixSectionRRD}.

\begin{theorem}
\label{TheoremOLSRDD} Let Assumptions \ref{assumption structural RDD} and  \ref{Relevance}  hold. Suppose also that Assumption \ref{AssumptionRRD} in
Section \ref{AppendixSectionRRD} in the Appendix hold. Then
\[
\sqrt{nh_{n}}(\hat{\beta}-\bar{\beta}_W)\rightarrow_{d}N\left(  0,\Sigma\right)
,
\]
 where  $\Sigma$ is given in Section \ref{AppendixSectionRRD} in the Appendix.
\end{theorem}

Remember from Proposition \ref{Identificationext} that if Assumption \ref{H} holds, then $\bar{\beta}_W=\beta.$  
The asymptotic variance of $\hat{\beta}$  can be consistently estimated using the standard 2SLS  variance estimator.

Theorem \ref{TheoremOLSRDD} can be immediately adapted to  the estimator in the parametric random coefficients model $\beta(W)=\theta_0+\theta_1\tilde{W}$ (Section \ref{sec:parametric}). In this case, all assumptions must hold with $\tilde{X}$ instead of $X$. Define $\tilde{\Sigma}$ identical to $\Sigma$ as in Section \ref{AppendixSectionRRD} in the Appendix, but   using $\tilde{X}$ instead of $X$. Then, the result is  $\sqrt{nh_{n}}(\hat{\theta}-\theta)\rightarrow_{d}N(  0,\tilde{\Sigma}),$ and $\tilde{\Sigma}$ can be consistently estimated using the standard 2SLS variance estimator. 

\begin{remark} \emph{\textbf{Some implementation considerations:}
  The identification results depend on the heterogeneity of the first stage discontinuities across the threshold for different values of $W.$ Therefore, what matters for identification is how many values $W$ assumes. It does not matter how many elements are in $W,$ and in fact $W$ could even be a single scalar variable. 
However, since for estimation  we will use a regression based technique, the heterogeneity of the first stage discontinuities is leveraged through the additional instruments which are obtained interacting the elements of $W$ with $D.$  Hence, $W$ needs to be a vector of at least $d-1$ elements. }

\emph{Thus, whichever form $W$ takes for identification, for estimation we need to transform it into a vector. The vector $W$ used in the estimation may  include any combination of the original variables, or transformations of those (indicators of some values of $W,$ interactions of elements of $W,$ higher order terms etc.) as long as the first stage vectors are sufficiently linearly independent, which is equivalent to saying that the instruments in the 2SLS are relevant. This can be verified in many ways, including visually in plots when $X$ is low dimensional, like in our application, through the \textit{ex post} consideration of the standard errors, or through a direct test (e.g. \cite{Shea}).}

\emph{In practice, therefore, all the considerations about the number of elements in $W$ in the identification sections should be understood as referring to the number of elements in the vector $W$ in the estimation plus one. Thus, if the $W$ used for estimation has $m$ elements, one should read the identification sections as if $q=m+1$. }

\emph{In the case of the linear model of equation \eqref{linear} in Section \ref{sec:parametric}, remember that $\tilde{W}$ is obtained from $W.$ Therefore, one must first decide what is the $W$ which will be used for estimation, and then decide $\tilde{W}$ respecting the dimension limits. Therefore, if $W$ has $m$ elements, then $\tilde{W}$ can have $c\leq (m+1)/d-1$ elements. }

\emph{It is important to note that although ${W}$ is used to build the instruments in our estimator,  there is never any requirement that $W$  be itself an instrumental variable, or even exogenous.} 
\end{remark}

\begin{remark} \emph{\textbf{Semiparametric estimation:} Theorem \ref{TheoremOLSRDD} assumes that  $E[Y|Z,W]=\alpha_{0Y}(Z)+\alpha_{1V}(Z)'W$ (equation \eqref{FSRDD} in Appendix \ref{AppendixSectionRRD}), which is in principle a semiparametric function of $W$. If the support of $W$ is discrete, say $\mathcal{W}=\{w_1,\dots,w_q\},$ as in our application, and we use in the estimation the vector  $W=(1(W=w_2),\dots,1(W=w_q)),'$ this is a fully nonparametric model. }

\emph{If the vector $W$ used for estimation does not saturate the support of $W,$ the  reduced-form equation is not fully nonparametric. However, one can define the vector $W$ used in the estimation to reflect a functional form that is as flexible as desired.}

\emph{If there is a desire for a fully nonparametric estimator in the continuous case, it can be done with minimal modifications to the estimator. Assume that  $E[Y|Z,W]\in L_2$, and simply define the vector $W$ used in the estimation  as  $(\varphi_1(W),\dots,\varphi_L(W))'$ where the $\varphi_l$ are elements of a sieve basis in the space of squared integrable functions of $W$. The rest of the estimation procedure remains unchanged. The  asymptotic results would need to be adapted to account for the fact that $L\rightarrow \infty$ as the sample increases (see e.g. \cite{Sieves}). It is expected that this estimator converges at the same rate as the semiparametric estimator, because the additional nonparametric estimation is compensated by the averaging in $W.$ Nevertheless, a formal proof of this is beyond the scope of this paper.}

\end{remark}
\section{An Application to the Estimation of the Effect of Insurance Coverage
on Health Care Utilization\label{application}}\label{application2}

We apply our approach to the problem of estimating the effects of insurance
coverage on health care utilization with a regression discontinuity design, as
in \citet{CardMedicare}, henceforth CDM. They exploit the fact that Medicare
eligibility varies discontinuously at age 65. Medicare eligibility may affect
health care utilization via two channels. First, it provides coverage to
people who were previously uninsured. Second, it provides more generous
coverage to people who are also insured by other
insurance policies. Let $Y$ be a measure of health care use (e.g., whether the
person did not get care for cost reasons last year.) The two treatment
variables  are an indicator of whether the person has any insurance coverage (i.e.,
one or more policies), $X_{1},$ and an indicator of whether the person has insurance coverage
by two or more policies, $X_{2}.$ The running variable $Z$ is defined to be
the age, measured in quarters of a year, relative to the threshold of 65 years
of age. The Medicare eligibility status $D$ is an indicator of whether the person is 65 years of age or older. We want to identify LATEs $\beta_{1}$ and $\beta_{2},$ which are respectively the local (to the threshold and to compliers) average treatment effects of $X_1$ and $X_2$ on $Y.$

Table \ref{SummaryStats} presents the summary statistics for the key variables
in our sample, obtained from the National Health Interview Survey (NHIS) from
1999 to 2003.
We consider three different outcome variables $Y$: whether the person
delayed care last year for cost reasons, whether the person did not get care
last year for cost reasons, and whether the person went to the hospital last
year.\footnote{There is another variable used by CDM: whether the person saw a doctor last year. This variable has many missing data throughout the sample
period, so we opted to drop it from our analysis. However, our conclusions do
not change when this variable is included in our study.}

\begin{table}[ht]
\caption{Summary Statistics}%
\label{SummaryStats}

\vspace{-.22in}
\begin{center}%
\begin{tabular}
[c]{l|c|ccc|ccc}%
\toprule Variable & All & \multicolumn{3}{c}{Non-Hispanic White} &
\multicolumn{3}{c}{Minority}\\
&  & HS & HS & Some & HS & HS & Some \\
&  & Dropout & Graduate & College & Dropout & Graduate & College \\[0.5ex]%
\hline
Delayed Care ($Y$) & 0.07 & 0.09 & 0.06 & 0.05 & 0.10 & 0.07 & 0.07\\
& (0.25) & (0.29) & (0.24) & (0.23) & (0.30) & (0.25) & (0.25)\\
Did Not Get Care ($Y$) & 0.05 & 0.07 & 0.04 & 0.03 & 0.09 & 0.06 & 0.05\\
& (0.22) & (0.25) & (0.20) & (0.18) & (0.29) & (0.23) & (0.22)\\
Went to Hospital ($Y$) & 0.13 & 0.17 & 0.12 & 0.12 & 0.14 & 0.12 & 0.12\\
& (0.34) & (0.38) & (0.33) & (0.32) & (0.35) & (0.33) & (0.32)\\
1+ Coverage ($X_{1}$) & 0.92 & 0.91 & 0.95 & 0.96 & 0.78 & 0.87 & 0.91\\
& (0.27) & (0.28) & (0.22) & (0.19) & (0.41) & (0.33) & (0.29)\\
2+ Coverage ($X_{2}$) & 0.33 & 0.44 & 0.38 & 0.32 & 0.24 & 0.23 & 0.24\\
& (0.47) & (0.50) & (0.49) & (0.47) & (0.43) & (0.42) & (0.42)\\
Medicare Eligible ($D$) & 0.42 & 0.55 & 0.45 & 0.37 & 0.46 & 0.36 & 0.34\\
& (0.49) & (0.50) & (0.50) & (0.48) & (0.50) & (0.48) & (0.47)\\
Observations & 63,165 & 8,337 & 16,037 & 21,352 & 8,293 & 4,302 & 4,844\\[1ex]%
 \bottomrule 
\end{tabular}
\end{center}

\vspace{-.05in}
{\footnotesize \begin{singlespace}Note: Source: NHIS 1999-2003. Standard deviations in
parentheses. ``HS Dropout" represents people who have less than a high school degree (DRP). ``HS Graduate" represents those who have exactly a high school degree (HS). ``Some College" represents those who have at least some college (COL). This sample reflects only people who are between 55 and 75 years old.\end{singlespace}}
\end{table}

The table shows that less educated people and minorities are more likely to delay or ration
care because of cost reasons. However, high school dropouts tend to go more to the hospital than their more educated counterparts, irrespective of their race. Additionally, non-Hispanic whites tend to carry more insurances than minorities with the same education level. Finally, people with more education are more likely to have some insurance than people with less education, irrespective of
the race, although heterogeneity along education is less pronounced than the heterogeneity found along race. 

\begin{figure}[H]
\caption{Age Distribution by Level of Education}%
\label{AgebyEduc}%

\vspace{-.25in}
\begin{center}
\includegraphics[scale=0.6]{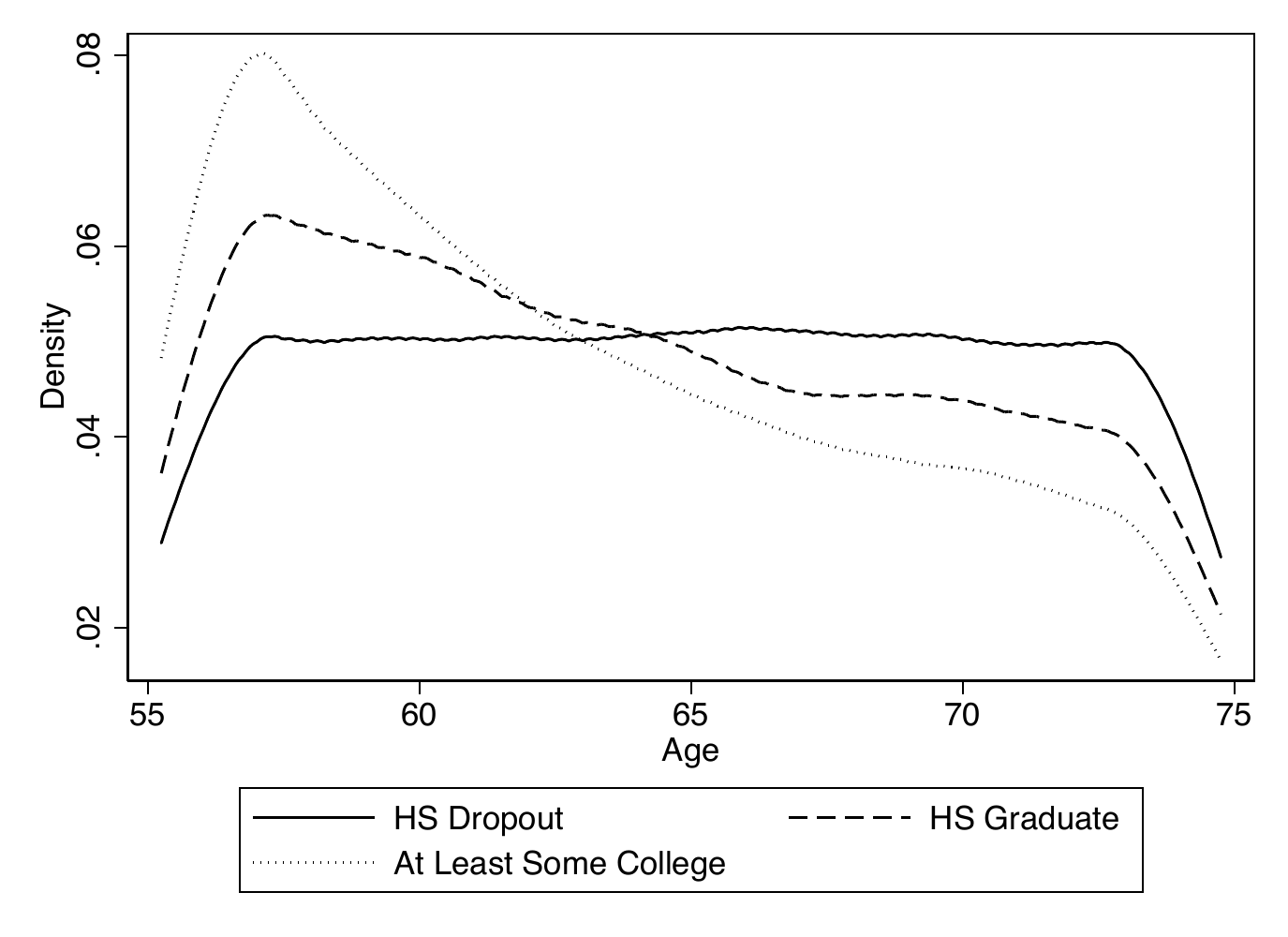}
\end{center}

\vspace{-.25in}
{\footnotesize \begin{singlespace}  Note: For each education level, this figure shows the kernel
density plot of the age distribution (measured in quarters of a year), using the 
Epanechnikov kernel and bandwidth equal to 1.\end{singlespace}}
\end{figure}

Interestingly, less educated whites are more likely to carry two or more
insurance policies than their more educated counterparts. This
counter-intuitive correlation is better understood in the context of age as an
important confounder. As seen in the second to last row of the table, people
with more education are less likely to be eligible to Medicare. Indeed, Figure
\ref{AgebyEduc} shows that people with less
 education tend to be older than
people with more education, so they are mechanically more likely to be eligible to
Medicare.

\begin{figure}[ht!]
\caption{Validity Plots}%
\label{Validity}%

\vspace{-.25in}
\begin{center}
\includegraphics[scale=0.6]{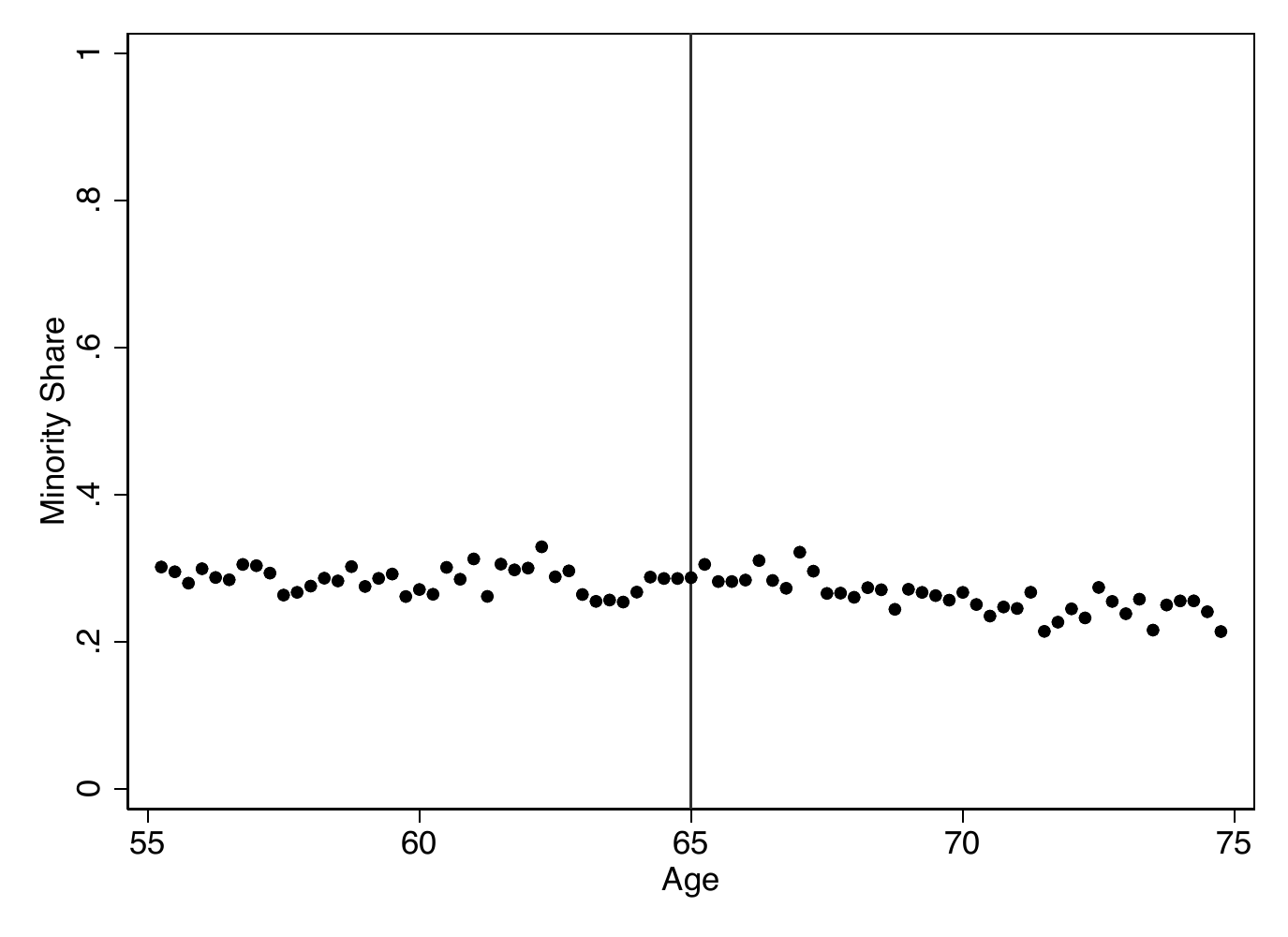}
\medspace\medspace \includegraphics[scale=0.6]{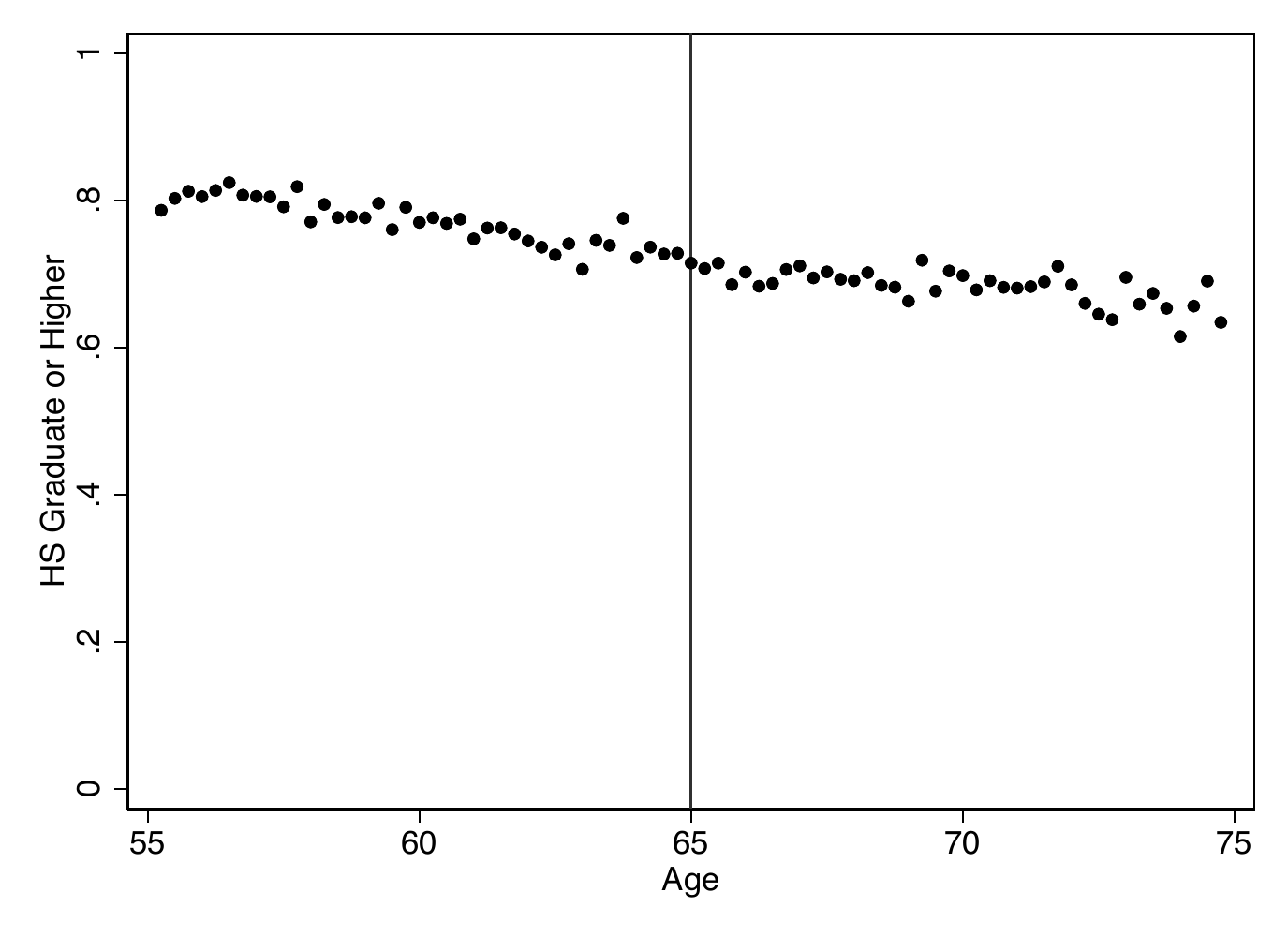}
\end{center}

\vspace{-.25in}
{\footnotesize \begin{singlespace} Note: The scatter plot shows the average of the variable
described in the vertical axis for each age level (measured in quarters of a
year).\end{singlespace}}
\end{figure}

 Following CDM, we use a regression discontinuity design to
circumvent this and other endogeneity concerns. To see how an RDD is reasonable in this context, Figure \ref{Validity} shows
that people just younger and just older than 65 years of age
are comparable in terms of race and education. Similarly to CDM, we find no
evidence of discontinuity at 65 years of age for a wide range of covariates.\footnote{Some of the potential confounders analyzed are the
person's income, work status and health status. See Section II.A in CDM for a
more detailed discussion of potential confounders in this context.}

The covariates available in the data include indicators for education (high school dropouts, DRP, high school graduates, HS, and those with at least some college, COL), race (non-Hispanic whites, WH, and minorities, MIN), gender, Hispanic status, region of the country (four indicators) and year of the sample (five indicators). Of these covariates, we must choose which ones to include in $W.$ The other variables not included in $W$ may be ignored, as in the standard RDD, or may be included in $C$ (see Section \ref{sec:homogeneityW1}) to improve efficiency, as in the RDD with covariates. 

To keep the application as close to CDM as possible, we  (1) include the additional covariates not included in $W$ as elements of $C$ (therefore, this application mirrors more  the RDD with covariates than the standard RDD);\footnote{The reported estimates are almost identical to the ones where we do not include these covariates in $C.$} (2)  report the results with bandwidth $h=10$ and uniform kernel;\footnote{We also implemented the approach with triangular kernel, and with bandwidths ranging from 5 to 10, obtaining very similar results.}  and (3) cluster standard errors by $Z.$\footnote{See \cite{Lee_Card} for a theoretical justification for clustering by values of the running variable in the RDD.}

In Table \ref{tab:main_results}, we report the TWLATE estimates $\bar{\beta}_{1,W}$ and $\bar{\beta}_{2,W}$ using our estimator (Section \ref{RDDestimation}) for different outcome variables $Y$ and different choices of $W.$ Under the exclusion restrictions $\beta_{1}(W)=\beta_{1}$ and $\beta_{2}(W)=\beta_{2}$ (Assumption \ref{H}, Section \ref{sec:exclusion}), the estimates are interpreted as the LATEs of $X_1$ and $X_2.$ When there are over-identifying restrictions, we test these exclusion restrictions with an over-identification test (the Sargan–Hansen's $J$-test), whose p-value is shown in brackets for each outcome $Y$ and each choice of $W.$
\begin{table}[ht!]
\protect\caption{Main Results \label{tab:main_results}}
\vspace{-0.2in}
\begin{center}
{\small \begin{tabularx}{.85\textwidth}{c|cc|cc|cc}

\toprule
 & \multicolumn{2}{c}{(I)} & \multicolumn{2}{c}{(II)} & \multicolumn{2}{c}{(III)} \\ & \multicolumn{2}{c}{$ W=$ Race} & \multicolumn{2}{c}{$ W=$ Race \& Educ} & \multicolumn{2}{c}{$ W=$ Race, Educ \& Other} \\ Outcome & $\hat{\beta}_{1}$ & $\hat{\beta}_{2}$ & $\hat{\beta}_{1}$ & $\hat{\beta}_{2}$ & $\hat{\beta}_{1}$ & $\hat{\beta}_{2}$ \tabularnewline
\midrule\addlinespace[1.5ex]
&-0.234**&0.008&-0.210**&0.004&-0.203**&0.001 \tabularnewline \addlinespace[.05in]
Delayed Care&(0.060)   &(0.013)   &(0.050)   &(0.010)   &(0.043)   &(0.009)    \tabularnewline \addlinespace[.05in]
&&&\multicolumn{2}{c}{[J-test p-value = 0.360]}&\multicolumn{2}{c}{[J-test p-value = 0.347]} \tabularnewline \addlinespace[.05in]
\midrule &-0.145**&-0.005&-0.132**&-0.007&-0.135**&-0.007 \tabularnewline \addlinespace[.05in]
Rationed Care&(0.053)   &(0.009)   &(0.043)   &(0.007)   &(0.038)   &(0.007)    \tabularnewline \addlinespace[.05in]
&&&\multicolumn{2}{c}{[J-test p-value = 0.458]}&\multicolumn{2}{c}{[J-test p-value = 0.133]} \tabularnewline \addlinespace[.05in]
\midrule &-0.087&0.038**&-0.109&0.041**&-0.084&0.035** \tabularnewline \addlinespace[.05in]
Went to Hospital&(0.086)   &(0.018)   &(0.081)   &(0.017)   &(0.074)   &(0.016)    \tabularnewline \addlinespace[.05in]
&&&\multicolumn{2}{c}{[J-test p-value = 0.925]}&\multicolumn{2}{c}{[J-test p-value = 0.516]} \tabularnewline \addlinespace[.05in]
\bottomrule \addlinespace[1.5ex]

\end{tabularx}
}
\end{center}
\vspace{-0.2in}
\footnotesize \begin{singlespace} Note: This table shows, for each horizontal panel represented by a different outcome variable $Y,$ 2SLS estimates of $\beta_{1}$ and $\beta_{2}$ under our multivariate RDD approach. These estimates were obtained under the exclusion restriction that $\beta_{1}(W)$ and $\beta_{2}(W)$ do not change with $W,$ where $W$ changes in the columns. Whenever possible, we also show the p-value of the test of this exclusion restriction via the Sargan-Hansen's $J$-test. In column (I), $W$ is an indicator for race. In column (II), we use indicators of the six groups formed by combinations of race (WH, MIN) and education (DRP, HS and COL). In column (III), we add indicators for gender, Hispanic status, region and year of the sample. In the three columns, all covariates available in the sample but not included in ${W}$ are included in $C$ (see Section \ref{sec:homogeneityW1} for a description of $C$) for efficiency, following CDM. These estimates were obtained from a local linear regression with uniform kernel and bandwidth $h=10$ (N=63,165), with standard errors clustered by $Z.$ *: significant at 10\%. **: significant at 5\%.\end{singlespace}
\end{table}

We start from the most parsimonious specification of $W.$ In column (I), we report the results for $W$ equal to an indicator of race. In this scenario we are just identified, so we are unable to test the exclusion restriction $\beta(\text{Race})=\beta.$ We find $\beta_1$ to be negative for ``Delayed Care'' and ``Ration Care," and  negative but insignificant for ``Went to Hospital." In contrast, we find that $\beta_2$ is close to zero for ``Delayed Care'' and ``Ration Care," and  positive for ``Went to Hospital.'' 

In column (II) we make $W$ more complex, including instead indicators of all combinations of race and education. This implies that the exclusion restriction must also include education ($\beta(\text{Race,Education})=\beta$), but at the same time we now have four over-identifying restrictions to test this assumption. The point estimates in all cases are very similar and are statistically the same as the ones from column (I). The standard errors are also uniformly smaller, now that we can exploit further heterogeneity in the first stages. The p-values from the J-tests suggest no evidence that the LATEs vary by either race or education, or both.

Finally, in column (III) we make $W$ even more complex by adding all the other remaining covariates in our dataset: gender, Hispanic status, region and year indicators. This greatly strengthens the exclusion requirements, by restricting the LATEs to not vary with these variables as well. At the same time, we provide more precise estimates, as we exploit further heterogeneity in the first stages. More importantly, now we have fourteen over-identifying restrictions to test the exclusion restrictions. As the p-values of the J-tests show, we now find some evidence that these exclusion restrictions might be invalid for ``rationed care.'' This leads us to choose the results from column (II) as our preferred ones, because there the exclusion restrictions seem to work for all outcome variables. However, it is interesting to note that the point estimates from column (III) are remarkably similar to the ones from column (II), suggesting that this over-identification test might have power to detect violations from the exclusion restrictions which do not yet bias our estimates too much.

Overall, we find that Medicare as the only insurance causes a reduction of about 15 percentage points in the delay or ration of health care for cost reasons, and Medicare as a second insurance causes an increase of about 4 percentage points in hospital visits. Our results confirm in a sharper way the intuition suggested by CDM that Medicare eligibility has generated different effects on health care utilization depending on whether the person had insurance before being eligible to Medicare.

\section{Conclusions}

\label{conclusion} This paper explores the identifiability of LATEs in the RDD with  multiple treatment values. We show that identification is generally impossible. In fact, even weighted averages of the LATEs of a specific treatment value are impossible to identify with useful weights. We characterize identification of weighted averages of all the LATEs and conclude that these averages can only be identified with weights that are proportional to the first-stage discontinuities.

If the first stage discontinuities for different values of  covariates are linearly independent, we can identify multivariate weighted averages of the LATEs of all treatment values which ``separate in expectation" the marginal treatment effects of one treatment value, in the sense that the weights average one for the LATEs of one treatment value, and average zero for the LATEs of all the other treatment values.

We then propose two strategies to reduce the space of the LATEs so as to achieve identification of the LATEs of a specific variable separately. The first strategy relies on homogeneity of the LATEs on some covariates, and the second strategy relies on a parametric specification. Both strategies can generate over-identification, which allows the practitioner to test whether the identification strategy adopted is valid.

We provide an estimator of the LATEs that can be programmed as a weighted Two Stage Least Squares regression on packaged software. The standard errors and test results obtained from packaged software can be used for inference.

We apply our method to the estimation of the effect of Medicare health insurance coverage for someone who was not previously insured separately from the effect of adding Medicare coverage for someone who already has another insurance. Our results show that having Medicare as the only insurance increases demand for health care, in the sense that it allows people to not delay or ration health care for costs reasons. However, going to the hospital is determined more by the intensity of coverage, so that having Medicare plus another insurance makes it more likely that someone would seek such expensive type of care.


\appendix

%

\section{Appendix: Estimation in the conditional case}\label{sec:heterogeneityW1}

In this section we adapt the estimator from Section \ref{sec:homogeneityW1} to allow us to estimate the conditional  TWLATEs $\bar{\beta}_{j,W,R}(R).$ Under Assumption \ref{Hc}, this estimator provides estimates of the conditional LATEs $\beta_j(R).$ The estimator is very similar to the estimator in Section \ref{sec:homogeneityW1}, with the following additional definitions:
\begin{enumerate}
    \item [5.] Let the support of $R$ be $\{r_{1},\dots,r_{q_R}\},$ then redefine $R=(1(R=r_{1}),\dots,1(R=r_{q_R}))'.$\footnote{This assumes that $R$ has a finite support. Indeed, since we have a finite number of instruments $D$ and $W\cdot D,$ this estimator can only be used whenever $q_R$ is finite. However, our theory allows the identification of the full function of TWLATEs $\bar{\beta}_{j,W}(r)$ or $\beta_j(r)$ when $R$ is a continuous variable. We suggest the following nonparametric estimator of $\beta_j(R)\in L_2$: redefine $R=(\varphi_1(R),\dots,\varphi_L(R))'$ where the $\varphi_l$ are elements of a sieve basis of $L_2$, then proceed with the estimator exactly as indicated.  Then $\hat{\beta}_j(R)=\hat{\beta}_{j1}\varphi_1(R)+\dots$ $\dots+\hat{\beta}_{jL}\varphi_L(R).$}
    \item [6.]  Redefine ${W}$ to be the vector of interactions of every element in $R$ with every element in $W.$ 
\item [7.] Define $\tilde{X}_{js}=X_j\cdot R_{s},$ which is the treatment for the observations such that $R=r_{s}.$ Define $\tilde{X}$ the  $d\!\cdot\! q_R\times 1$ vector of all $\tilde{X}_{js}.$

\end{enumerate}
We formalize our approach as a weighted 2SLS regression with first-stage regressions
\begin{equation*}
    \tilde{X}_{ijs}=\pi_{0js}D_i+\pi_{1js}'{W}_i\cdot D_i+\alpha_{js}'C_i+\varepsilon_{ijs}^x, \quad j=1,\dots,d,\;\; s=1,\dots,q_1,
\end{equation*}
second stage regression
\begin{equation*}
    \hspace{-2.5cm}{Y}_{i}=\beta'\widehat{\tilde{X}}_i+\eta'C_i+\varepsilon_{i}^y,
\end{equation*}
and weights equal to $k_{h_{n}}(Z_i).$ In other words, we propose running a weighted  2SLS regression of $Y$ on $\tilde{X}$ using $D$ and $D\cdot {W}$ as IVs,  $C$ as exogenous controls, and the $k_{h_n}$ as weights.\footnote{In Stata, the code is ``ivregress 2sls Y C (X RX = D ${\text{W}}$D) [k]".}  Standard errors obtained directly  from packaged software can be used for inference.

In general, the estimated coefficient of $\tilde{X}_{js},$ $\hat{\beta}_{js},$ estimates the conditional TWLATE  $\bar{\beta}_{j,W,R}(r_{s}).$ Under  the assumption $\beta(W,R)=\beta(R)$ (Section \ref{sec:exclusionR}),  $\hat{\beta}_{js}$ estimates the conditional LATEs  ${\beta}_j(r_{s}).$

Given the way $R$ is redefined, it can be shown that the estimator in this section is equivalent to a series of $q_R$ estimators from Section \ref{sec:homogeneityW1}, each restricted only to data such that $R=r_{s},$ for $s=1,\dots,q_R.$ Because of this, we refer the reader to that section for details about implementation.

\section{Appendix: Proofs}

\subsection{Identification Results}

\noindent\textbf{Proof of Proposition \ref{Lack}}:\ The identified set for
$\beta(w)$ is
\begin{equation}
\{\beta_{0}(w)+n(w):\text{where }\delta_{Y}(w)=\beta_{0}^{\prime}(w)\delta
_{X}(w)\text{ and }n^{\prime}(w)\delta_{X}(w)=0\}, \label{A1}%
\end{equation}
where $\beta_{0}(w)$ is the true value that generated the data. Since there are an infinite number of $n$'s satisfying the conditions in \eqref{A1}, the result follows.\bigskip

\noindent\textbf{Proof of Proposition \ref{Local}}:\ If there exists $w%
\in\mathcal{W}$ and $j$ such that $\delta_{X_{-j}}(w)=0$ and
$\delta_{X_{j}}(w)\neq0.$ Then, $\delta_{Y}(w)=\beta_{j}(w%
)\delta_{X_{j}}(w)$ and hence $\beta_{j}(w)$ is identified as
$\delta_{Y}(w)/\delta_{X_{j}}(w).$ The necessity follows from the next proposition applied to an $\omega$ equals to a Delta Dirac measure. \bigskip

\noindent\textbf{Proof of Proposition \ref{WATE}}:\ Without loss of generality
take $j=1$. From (\ref{A1}) a necessary and sufficient condition for
identification of $E[\omega_{1}(W)\beta_{1}(W)]$ is $E[\omega_{1}%
(W)n_{1}(W)]=0$ for all $n^{\prime}(w)\delta_{X}(w)=0$, where $n_{1}(w)$ denotes the first component of $n(w)$. Note that a basis for
$\mathbb{R}^{d}$ can be constructed with first vector $\delta_{X}(w)$ and
where the first elements for the rest of the vectors in the basis are
$\delta_{X_{j}}(w)$ for $j=2,...,d.$ For example, $v_{2}(w)=(\delta_{X_{2}%
}(w),-\delta_{X_{1}}(w),0,...,0)^{\prime}$ satisfies the orthogonality
condition, and similarly we construct $v_{3}(w),...,v_{d}(w)=(\delta_{X_{d}%
}(w),0,...,0,-\delta_{X_{1}}(w))^{\prime}.$ Any linear combination of these
vectors satisfy the orthogonality condition. Take $n(w)=\alpha(w)v_{2}%
(w)$, for a scalar $\alpha(w).$ Then, $n^{\prime}(w)\delta_{X}(w)=0$, and
for identification we need $E[\omega_{1}(W)\alpha(W)\delta_{X_{2}}(W)]=0$.
Taking $\alpha(w)=\omega_{1}(w)\delta_{X_{2}}(w)$ conclude $\omega
_{1}(w)\delta_{X_{2}}(w)=0$ a.s. The same argument applied to $v_{j}$ yields
$\omega_{1}(w)\delta_{X_{j}}(w)=0$ a.s. for $j=3,..,d$. For the sufficiency
note that by the previous arguments the first element $n_{1}(w)$ of a vector
satisfying $n^{\prime}(w)\delta_{X}(w)=0$ can be written as linear combination
of the $\delta_{X_{j}}(w)$, for $j=2,...,d$, and thus $E[\omega_{1}(W)n_{1}(W)]=0.$
\bigskip

\noindent\textbf{Proof of Proposition \ref{MWATE}}:\ From (\ref{A1}) a
necessary and sufficient condition for identification of the TWLATE $E[\omega(W)^{\prime
}\beta(W)]$ is $E[\omega(W)^{\prime}n(W)]=0$ for all $n^{\prime}(w)\delta
_{X}(w)=0.$ The sufficiency is then clear, since $E[a(W)\delta_{X}(W)^{\prime
}n(W)]=0$. For the necessity, use that any $n(W)$ satisfying $n^{\prime
}(w)\delta_{X}(w)=0$ can be written as a linear combination of the vectors
$v_{j}(w)$ for $j=2,...,d$, say $n(w)=n_{2}(w)v_{2}(w)+\cdots+n_{d}%
(w)v_{d}(w)$. Then, $0=E[\omega(W)^{\prime}n(W)]=E[\omega_{2}(W)^{\prime}%
n_{2}(W)]+\cdots+E[\omega_{d}(W)^{\prime}n_{d}(W)]$, where $\omega
_{j}(w)=\omega(w)^{\prime}v_{j}(w)$. Taking $n_{j}(w)=\omega_{j}(w)$ we
conclude $\omega_{j}(w)=0$ a.s. for $j=2,...,d$ and hence $\omega(w)$ must be
proportional to $\delta_{X}(w)$, which shows the necessity.\bigskip

\noindent\textbf{Proof of Proposition \ref{cor:total}}:\ For the necessity, note that if Assumption \ref{Relevance} does not hold then there is a non-trivial linear combination of the first stages $\delta_X(W)'v=0$ a.s., and thus $E[a(W)\delta_{X}(W)']=I$ cannot hold (because $E[v'a(W)\delta_{X}(W)'v]=0$). The sufficiency is shown right after the proposition.\bigskip

\comment{
\bigskip
\noindent\textbf{Proof of Proposition \ref{MWATERDD}}:\ Using the definition
of conditional expectation, and dominated convergence (which holds by
Assumption \ref{assumption structural RDD}(iv)) it follows
\begin{align*}
\beta_{j}  & =\lim_{e\downarrow0}E[\beta_{ij}|\Delta_{ij}(e)=1]\\
& =\lim_{e\downarrow0}\frac{E[\beta_{ij}\Delta_{ij}(e)]}{E[\Delta_{ij}(e)]}\\
& =\lim_{e\downarrow0}\frac{E[E[\beta_{ij}|W,\Delta_{ij}(e)]\Delta_{ij}%
(e)]}{E[\Delta_{ij}(e)]}\\
& =\frac{\lim_{e\downarrow0}E[E[\beta_{ij}|W,\Delta_{ij}(e)]\Delta_{ij}%
(e)]}{\lim_{e\downarrow0}E[\Delta_{ij}(e)]}\\
& =\frac{E[\lim_{e\downarrow0}E[\beta_{ij}|W,\Delta_{ij}(e)]\lim
_{e\downarrow0}\Delta_{ij}(e)]}{\lim_{e\downarrow0}E[\Delta_{ij}(e)]}\\
& =\frac{E[\beta_{j}(W)\lim_{e\downarrow0}\Delta_{ij}(e)]}{\lim_{e\downarrow
0}E[\Delta_{ij}(e)]}\\
& =\lim_{e\downarrow0}\frac{E[\beta_{j}(W)\Delta_{ij}(e)]}{E[\Delta_{ij}%
(e)]}\\
& =\lim_{e\downarrow0}E[\beta_{j}(W)|\Delta_{ij}(e)=1].
\end{align*}
Thus, again by dominated convergence
\begin{align*}
\beta_{RDD}  & =\sum_{j=1}^{d}\beta_{j}\pi_{j}\\
& =\sum_{j=1}^{d}\lim_{e\downarrow0}E[\beta_{j}(W)\pi_{j}|\Delta_{ij}(e)=1]\\
& =\sum_{j=1}^{d}\lim_{e\downarrow0}\frac{E[\beta_{j}(W)\pi_{j}E[\Delta
_{ij}(e)|W]]}{E[\Delta_{ij}(e)]}\\
& =\sum_{j=1}^{d}\frac{E[\beta_{j}(W)\pi_{j}\delta_{X_{j}}(W)]}{\delta_{X_{j}%
}}\\
& =\sum_{j=1}^{d}\frac{E[\beta_{j}(W)\delta_{X_{j}}(W)]}{\delta_{T}}.
\end{align*}
}

\bigskip

\noindent\textbf{Proof of Proposition \ref{Identificationext}}:\ We prove the more general conditional version. Taking limits as in the main
text we obtain the equation%
\[
\delta_{Y}(W,R)=\delta_{X}^{\prime}(W,R)\beta(R).
\]
Multiplying by $\delta_{X}(W,R)$ both sides, and taking conditional means on
$R,$ we arrive at
\[
E[\delta_{X}(W,R)\delta_{Y}(W,R)|R]=E[\delta_{X}(W,R)\delta_{X}^{\prime
}(W,R)|R]\beta(R).
\]
This and the generalized relevance condition yield identification, i.e.
\[
\beta(R)=\left(  E[\delta_{X}(W,R)\delta_{X}^{\prime}(W,R)|R]\right)
^{-1}E[\delta_{X}(W,R)\delta_{Y}(W,R)|R].
\]
For the necessity part, we suppose the generalized relevance condition does
not hold. This means there exists a non-trivial measurable function
$\lambda(R)$ such that, a.s.,
\[
\lambda(R)^{\prime}E[\delta_{X}(W,R)\delta_{X}^{\prime}(W,R)|R%
]\lambda(R)=0.
\]
Hence,%
\[
E\left[\left(  \lambda(R)^{\prime}\delta_{X}(W,R)\right)  ^{2}\right]=0,
\]
and thus
\begin{equation*}
\lambda(R)^{\prime}\delta_{X}(W,R)=0\text{ a.s}. 
\end{equation*}
Let $\beta_{0}(R)$ denote the true value that generated the data. Then, $\tilde{\beta}(R)=\beta_{0}(R)+\lambda(R)$ is observationally equivalent to $\beta_{0}(R)$, and thus $\beta_{0}(\cdot)$ is not identified.

\subsection{Asymptotic Theory}

\label{AppendixSectionRRD}

In this section we establish the asymptotic normality of the proposed
estimator $\hat{\beta}.$ Write $W_{i}=(W_{1i},...,W_{mi})^{\prime}$. Without
loss of generality assume hereinafter that $z_{0}=0.$ We introduce some
further notation and assumptions. Let $\varepsilon_{V_{i}}=V_{i}-E[V_{i}%
|Z_{i},W_{i}]$ denote the regression errors for $V=Y$ and $V=X.$ Define
$\zeta_{i}=\varepsilon_{Y_{i}}-\bar{\beta}_{W}^{\prime}\varepsilon_{X_{i}}.$
Assume, for $V_{i}=Y_{i}$ and $V_{i}=X_{ij},$ $j=1,...,d,$%
\begin{equation}
E[V_{i}|Z_{i},W_{i}]=\alpha_{0V}(Z_{i})+\alpha_{1V}(Z_{i})W_{i1}%
+\cdots++\alpha_{mV}(Z_{i})W_{im}.\label{FSRDD}%
\end{equation}
\noindent For notational convenience, in the proofs we use the notation
$H_{i}=Y_{i}(0)$ for the random intercept, rather than $\alpha_{i}=Y_{i}(0)$
(since $\alpha$ is used several times for generic coefficients). In the proofs
we use the generic notation, for a generic $V_{i},$ $j,g=1,...,m,$
\begin{align*}
\mu_{jg}^{V}(z) &  =E[W_{ij}W_{ig}V_{i}|Z_{i}=z]\\
\sigma_{jg}^{2,V}(z) &  =E[W_{ij}^{2}W_{ig}^{2}V_{i}^{2}|Z_{i}=z]\\
\kappa_{jg}^{V}(z) &  =E[W_{ij}^{3}W_{ig}^{3}V_{i}^{3}|Z=z].
\end{align*}
When $V$ is identically $1,$ we drop the reference to $V$ above.

We investigate the asymptotic properties of $\hat{\beta}$ under the following
assumptions, which parallel those of \cite{HTV2001}:\bigskip

\begin{assumption}
\label{AssumptionRRD} Suppose that

\begin{enumerate}
\item The sample $\{\chi_{i}\}_{i=1}^{n}$ is an iid sample, where $\chi
_{i}=(Y_{i},X_{i}^{\prime},W_{i}^{\prime},Z_{i})^{\prime}.$

\item The density of $Z,$ $f(\cdot),$ is continuous and bounded near $z=0.$ It
is also bounded away from zero near $z=0.$ The matrices $\Gamma$ and
$\Delta_{X}$, defined below, are positive definite and of full column rank, respectively.

\item The kernel $k$ is continuous, symmetric and nonnegative-valued with
compact support.

\item The functions $\mu_{jg}(z),$ $\mu_{jg}^{X}(z),$ $\mu_{jg}^{Y}(z),$
$\sigma_{jg}^{2}(z),$ $\sigma_{jg}^{2,X}(z),$ $\sigma_{jg}^{2,Y}(z),$
$\sigma_{jg}^{2,\zeta}(z),$ and $\kappa_{jg}^{\zeta}(z)$ are uniformly bounded
near $z=0,$ with well-defined and finite left and right limits to $z=0,$ for
$j,g=1,...,m.$

\item The bandwidth satisfies $nh_{n}^{5}\rightarrow0.$

\item For $V_{i}=Y_{i}$ and $V_{i}=X_{il}$ $l=1,...,d$: (i) let equation
\eqref{FSRDD} hold; for $j=0,...,m:$ (ii) for $z>0$ or $z<0,$ $\alpha_{jV}(z)$
is twice continuously differentiable; (iii) there exists some $M>0$ such that
$\dot{\alpha}_{jV}^{+}(Z)=\lim_{u\downarrow Z}\partial\alpha_{jV}(u)/\partial
u$ and $\ddot{\alpha}_{jV}^{+}(Z)=\lim_{u\downarrow Z}\partial^{2}\alpha
_{jV}(u)/\partial u^{2}$ are uniformly bounded on $(0,M].$ Similarly,
$\dot{\alpha}_{jV}^{-}(Z)=\lim_{u\uparrow Z}\partial\alpha_{jV}(u)/\partial u$
and $\ddot{\alpha}_{jV}^{-}(Z)=\lim_{u\uparrow Z}\partial^{2}\alpha
_{jV}(u)/\partial u^{2}$ are uniformly bounded on $[-M,0).$\medskip
\end{enumerate}
\end{assumption}

For a measurable function of the data $\varphi(\chi_{i}),$ define the local
linear sample mean
\[
\hat{E}[\varphi(\chi_{i})]=\frac{1}{nh_{n}}\sum_{i=1}^{n}\varphi(\chi
_{i})k_{ih_{n}}.
\]
Define $S_{+i}=D_{i}\cdot(1,Z_{i},W_{i}^{\prime},W_{i}^{\prime}Z_{i})^{\prime
}$, $S_{-i}=(1-D_{i})\cdot(1,Z_{i},W_{i}^{\prime},W_{i}^{\prime}Z_{i}%
)^{\prime}$, and $S_{i}=(S_{+i}^{\prime},S_{-i}^{\prime})^{\prime}$. Recall
$C=(1,{W}^{\prime},Z,Z\cdot D,Z\cdot{W}^{\prime},Z\cdot D\cdot{W}^{\prime
})^{\prime}.$ Note the spanning of the exogenous variables $(D_{i}%
,W_{i}^{\prime}D_{i},C_{i})$ is the same as that of $S_{i}$, and thus the
first stage is the regression of $X_{i}$ on $S_{i}$. Let $S_{in}$ and $C_{in}$
be defined the same as $S_{i}$ and $C_{i}$ but with $Z_{i}$ replaced by
$Z_{i}/h_{n}.$ Define $\tilde{X}_{i}=(X_{i}^{\prime},C_{i}^{\prime})^{\prime
},$ $\tilde{X}_{in}=(X_{i}^{\prime},C_{in}^{\prime})^{\prime},$ and
$\tau=(\bar{\beta}_{W}^{\prime},\eta^{\prime})^{\prime},$ where $\eta
=(\eta_{1},\eta_{2}^{\prime},\eta_{3},\eta_{4},\eta_{5}^{\prime},\eta
_{6}^{\prime})^{\prime}$ has the same dimension as $C_{i}$ and is such that%
\[
\eta^{\prime}C_{i}=\eta_{1}+\eta_{2}^{\prime}{W}_{i}+\eta_{3}Z_{i}+\eta
_{4}D_{i}Z_{i}+\eta_{5}^{\prime}{W}_{i}Z_{i}+\eta_{6}^{\prime}W_{i}D_{i}%
Z_{i}.
\]
Define $\tau_{n}=(\bar{\beta}_{W}^{\prime},\eta_{1},\eta_{2}^{\prime},\eta
_{3}h_{n},h_{n}\eta_{4},h_{n}\eta_{5}^{\prime},h_{n}\eta_{6})^{\prime},$ so
that $\tilde{X}_{i}^{\prime}\tau=\tilde{X}_{in}^{\prime}\tau_{n}.$ With this
notation in place, the 2SLS is the first $d\times1$ subcomponent of%
\begin{align*}
\hat{\tau}_{n}  &  =\left(  \hat{E}\left[  \tilde{X}_{in}S_{in}^{\prime
}\right]  \left(  \hat{E}\left[  S_{in}S_{in}^{\prime}\right]  \right)
^{-1}\hat{E}\left[  S_{in}\tilde{X}_{in}^{\prime}\right]  \right)  ^{-1}%
\hat{E}\left[  \tilde{X}_{in}S_{in}^{\prime}\right]  \left(  \hat{E}\left[
S_{in}S_{in}^{\prime}\right]  \right)  ^{-1}\hat{E}\left[  S_{in}Y_{i}\right]
\\
&  =\tau_{n}+\left(  \hat{E}\left[  \tilde{X}_{in}S_{in}^{\prime}\right]
\left(  \hat{E}\left[  S_{in}S_{in}^{\prime}\right]  \right)  ^{-1}\hat
{E}\left[  S_{in}\tilde{X}_{in}\right]  \right)  ^{-1}\hat{E}\left[  \tilde
{X}_{in}S_{in}^{\prime}\right]  \left(  \hat{E}\left[  S_{in}S_{in}^{\prime
}\right]  \right)  ^{-1}\hat{E}\left[  S_{in}U_{i}\right]  ,
\end{align*}
where $U_{i}=Y_{i}-\tilde{X}_{in}^{\prime}\tau_{n}=Y_{i}-\tilde{X}_{i}%
^{\prime}\tau.$

We show in Lemma \ref{MainCLT} that
\begin{equation}
\sqrt{nh_{n}}\left(  \hat{\tau}_{n}-\tau_{n}\right)  \rightarrow_{d}%
N(0,\Omega), \label{step2}%
\end{equation}
and provide an expression for $\Omega.$ The asymptotic normality for
$\sqrt{nh_{n}}\left(  \hat{\beta}_{n}-\bar{\beta}_{W}\right)  $ then follows
as%
\[
\sqrt{nh_{n}}\left(  \hat{\beta}_{n}-\bar{\beta}_{W}\right)  \rightarrow
_{d}N(0,\Sigma),
\]
where $\Sigma=e_{0}^{\prime}\Omega e_{0},$ and $e_{0}$ is a selection matrix
that selects the first $d$ elements of $\hat{\tau}_{n}$.

We introduce some notation that will be used throughout,%
\[
\gamma_{l}=\int_{0}^{\infty}u^{l}k(u)du,
\]%
\[
\mu_{jg}^{+,V}=\lim_{z\downarrow0}E[W_{ij}W_{ig}V_{i}|Z_{i}=z]\qquad\mu
_{j\rho}^{-,V}=\lim_{z\uparrow0}E[W_{ij}W_{ig}V_{i}|Z_{i}=z],
\]%
\[
k_{ih_{n}}^{+}=k(Z_{i}/h_{n})1(Z_{i}\geq0)\qquad k_{ih_{n}}^{-}=k(Z_{i}%
/h_{n})1(Z_{i}<0),
\]%
\[
b_{ljg}^{+}=\gamma_{l}\mu_{jg}^{+},\text{ }b_{ljg}^{-}=\gamma_{l}\mu_{jg}%
^{-},\text{ }b_{ljg}^{+,X}=\gamma_{l}\mu_{jg}^{+,X},\text{ }b_{ljg}%
^{-,X}=\gamma_{l}\mu_{jg}^{-,X}.
\]
We also use the notation that dropping an index means dropping the variable
associated to the index, so for example, $\mu_{j}^{+,V}=\lim_{z\downarrow
0}E[W_{ij}V_{i}|Z_{i}=z]$ and $\mu^{+,V}=\lim_{z\downarrow0}E[V_{i}|Z_{i}=z],$
and when $V$ is dropped from the notation it means is substituted by $1.$ The
notation $\mu_{W}^{+,V}=(\mu_{1}^{+,V},...,\mu_{m}^{+,V})$ means a row vector
with the same dimension as $W,$ and $\mu_{\otimes}^{+,V}$ is the $m\times m$
matrix with elements $\mu_{jg}^{+,V}.$ Similarly, we denote $b_{lW}%
^{+,X}=\gamma_{l}\mu_{W}^{+,X}$ and $b_{l\otimes}^{+,X}=\gamma_{l}\mu
_{\otimes}^{+,X},$ and likewise for other quantities.

\begin{lemma}
\label{Deltax} Under Assumption \ref{AssumptionRRD} 1-5,
\[
\hat{E}\left[  \tilde{X}_{in}S_{in}^{\prime}\right]  \rightarrow_{p}\Delta
_{X},
\]
where
\[
\Delta_{X}=f(0)\left[
\begin{array}
[c]{cccccccc}%
b_{00}^{+,X} & b_{10}^{+,X} & b_{0W}^{+,X} & b_{1W}^{+,X} & b_{00}^{-,X} &
b_{10}^{-,X} & b_{0W}^{-,X} & b_{1W}^{-,X}\\
b_{00}^{+} & b_{10}^{+} & b_{0W}^{+} & b_{1W}^{+} & b_{00}^{-} & b_{10}^{-} &
b_{0W}^{-} & b_{1W}^{-}\\
b_{0W}^{+\prime} & b_{1W}^{+\prime} & b_{0\otimes}^{+} & b_{1\otimes}^{+} &
b_{0W}^{-\prime} & b_{1W}^{-\prime} & b_{0\otimes}^{-} & b_{1\otimes}^{-}\\
b_{1}^{+} & b_{2}^{+} & b_{1W}^{+} & b_{2W}^{+} & b_{1}^{-} & b_{2}^{-} &
b_{1W}^{-} & b_{2W}^{-}\\
b_{1}^{+} & b_{2}^{+} & b_{1W}^{+} & b_{2W}^{+} & 0 & 0 & 0 & 0\\
b_{1W}^{+\prime} & b_{2W}^{+\prime} & b_{1\otimes}^{+} & b_{2\otimes}^{+} &
b_{1W}^{-\prime} & b_{2W}^{-\prime} & b_{1\otimes}^{-} & b_{2\otimes}^{-}\\
b_{1W}^{+\prime} & b_{2W}^{+\prime} & b_{1\otimes}^{+} & b_{2\otimes}^{+} &
0 & 0 & 0 & 0
\end{array}
\right]  .
\]

\begin{proof}
Let
\[
\tau_{ljg}^{+}=\frac{1}{nh_{n}}\sum_{i=1}^{n}\left(  \frac{Z_{i}}{h_{n}%
}\right)  ^{l}W_{ij}W_{ig}k_{ih_{n}}^{+},\qquad l,\rho=0,1,2,j=1,...,m.
\]
Then, by the change of variables $u=Z/h_{n},$%
\begin{align*}
E[\tau_{ljg}^{+}]  &  =h_{n}^{-1}E\left[  \left(  \frac{Z_{i}}{h_{n}}\right)
^{l}W_{ij}W_{ig}k_{ih_{n}}^{+}\right] \\
&  =\int_{0}^{\infty}u^{l}k(u)\mu_{jg}(uh_{n})f(uh_{n})du\\
&  =f(0)\gamma_{l}\mu_{jg}^{+}+o(1),
\end{align*}
where $\mu_{jg}(z)=E[W_{ij}W_{ig}|Z_{i}=z]$ and the convergence follows by the
Dominated Convergence theorem. As for the variance%
\begin{align*}
Var(\tau_{ljg}^{+})  &  \leq\left(  nh_{n}^{2}\right)  ^{-1}E\left[  \left(
\frac{Z_{i}}{h_{n}}\right)  ^{2l}W_{ij}^{2}W_{ig}^{2}k_{ih_{n}}^{+2}\right] \\
&  =\left(  nh_{n}\right)  ^{-1}\int_{0}^{\infty}u^{2l}k^{2}(u)\sigma_{jg}%
^{2}(uh_{n})f(uh_{n})du\\
&  =o(1),
\end{align*}
again by the Dominated Convergence theorem.

Similarly, define
\[
\tau_{ljg}^{+,X}=\frac{1}{nh_{n}}\sum_{i=1}^{n}\left(  \frac{Z_{i}}{h_{n}%
}\right)  ^{l}W_{ij}W_{ig}X_{i}k_{ih_{n}}^{+},\qquad l=0,1,2,j,g=1,...,m.
\]
Then, by the change of variables $u=Z/h_{n},$%
\begin{align*}
E[\tau_{ljg}^{+,X}]  &  =h_{n}^{-1}E\left[  \left(  \frac{Z_{i}}{h_{n}%
}\right)  ^{l}W_{ij}W_{ig}X_{i}k_{ih_{n}}^{+}\right] \\
&  =\int_{0}^{\infty}u^{l}k(u)\mu_{jg}^{X}(uh_{n})f(uh_{n})du\\
&  =f(0)\gamma_{l}\mu_{jg}^{+,X}+o(1),
\end{align*}
where $\mu_{jg}^{X}(z)=E[W_{ij}W_{ig}X_{i}|Z_{i}=z],$ by the Dominated
Convergence theorem. As for the variance, when $d=1,$%
\begin{align*}
Var(\tau_{ljg}^{+,X})  &  \leq\left(  nh_{n}^{2}\right)  ^{-1}E\left[  \left(
\frac{Z_{i}}{h_{n}}\right)  ^{2l}W_{ij}^{2}W_{ig}^{2}X_{i}^{2}k_{ih_{n}}%
^{+2}\right] \\
&  =\left(  nh_{n}\right)  ^{-1}\int_{0}^{\infty}u^{2l}k^{2}(u)\sigma
_{jg}^{2,X}(uh_{n})f(uh_{n})du\\
&  =o(1),
\end{align*}
again by the Dominated Convergence theorem. The same argument holds when
$d>1.$ The proof for $\tau_{ljg}^{-}$ and $\tau_{ljg}^{-,X},$ which replace
$k_{ih_{n}}^{+}$ by $k_{ih_{n}}^{-},$ is analogous, and hence omitted.
\end{proof}
\end{lemma}

\begin{lemma}
\label{S} Under Assumption \ref{AssumptionRRD} 1-5,%
\[
\hat{E}\left[  S_{in}S_{in}^{\prime}\right]  \rightarrow_{p}\Gamma
\]
where
\[
\Gamma=f(0)\left[
\begin{array}
[c]{cccccccc}%
b_{0}^{+} & b_{1}^{+} & b_{0W}^{+} & b_{1W}^{+} & 0 & 0 & 0 & 0\\
b_{1}^{+} & b_{2}^{+} & b_{1W}^{+} & b_{2W}^{+} & 0 & 0 & 0 & 0\\
b_{0W}^{+\prime} & b_{1W}^{+\prime} & b_{0\otimes}^{+} & b_{1\otimes}^{+} &
0 & 0 & 0 & 0\\
b_{1W}^{+\prime} & b_{2W}^{+\prime} & b_{1\otimes}^{+} & b_{2\otimes}^{+} &
0 & 0 & 0 & 0\\
0 & 0 & 0 & 0 & b_{0}^{-} & b_{1}^{-} & b_{0W}^{-} & b_{1W}^{-}\\
0 & 0 & 0 & 0 & b_{1}^{-} & b_{2}^{-} & b_{1W}^{-} & b_{2W}^{-}\\
0 & 0 & 0 & 0 & b_{0W}^{-\prime} & b_{1W}^{-\prime} & b_{0\otimes}^{-} &
b_{1\otimes}^{-}\\
0 & 0 & 0 & 0 & b_{1W}^{-\prime} & b_{2W}^{-\prime} & b_{1\otimes}^{-} &
b_{2\otimes}^{-}%
\end{array}
\right]  .
\]

\begin{proof}
The proof is analogous to that of Lemma \ref{Deltax} and hence is omitted.
\bigskip
\end{proof}
\end{lemma}

\begin{lemma}
\label{Deltay} Under Assumption \ref{AssumptionRRD} 1-5,%
\[
\hat{E}\left[  S_{in}Y_{i}\right]  \rightarrow_{p}\Delta_{Y},
\]
where $\Delta_{Y}=f(0)(b_{00}^{+,Y},b_{10}^{+,Y},b_{0W}^{+,Y},b_{1W}%
^{+,Y},b_{00}^{-,Y},b_{10}^{-,Y},b_{0W}^{-,Y},b_{1W}^{-,Y})^{\prime}.$

\begin{proof}
The proof is analogous to that of Lemma \ref{Deltax} and hence is omitted.
\end{proof}
\end{lemma}

\begin{lemma}
\label{consistency} Assume \ref{assumption structural RDD}, \ref{Relevance}
and \ref{AssumptionRRD}. Then,
\[
\Delta_{Y}=\Delta_{X}^{\prime}\tau_{0},
\]
where $\tau_{0}=(\bar{\beta}_{W}^{\prime},\eta_{1},\eta_{2}^{\prime
},0,0,0^{\prime},0^{\prime})^{\prime}.$

\begin{proof}
By Assumption \ref{AssumptionRRD} 6, we can write%
\[
E[H_{i}|W_{i},Z_{i}]=\alpha_{0H}(Z_{i})+\alpha_{1H}(Z_{i})W_{1i}+\cdots
+\alpha_{mH}(Z_{i})W_{mi}.
\]
Take for example $b_{00}^{+,Y}\!.$ By substitution of the random coefficients
representation and the homogeneity and identifying assumptions%
\[
b_{00}^{+,Y}=\bar{\beta}_{W}b_{00}^{+,X}+\eta_{1}b_{00}^{+}+\eta_{2}^{\prime
}b_{0W}^{+},
\]
where%
\[
\eta_{1}=\alpha_{0H}(0)
\]
and%
\[
\eta_{2}=(\alpha_{1H}(0),...,\alpha_{mH}(0))^{\prime}.
\]
The same argument applied to other components of $\Delta_{Y},$ lead to the
desired emuality $\Delta_{Y}=\Delta_{X}^{\prime}\tau_{0}.$
\end{proof}
\end{lemma}

The previous Lemmas show the consistency of $\hat{\beta},$ since
\[
\hat{\tau}_{n}\rightarrow_{p}\left(  \Delta_{X}\Gamma^{-1}\Delta_{X}^{\prime
}\right)  ^{-1}\Delta_{X}\Gamma^{-1}\Delta_{Y}=\tau_{0}%
\]
We prove several Lemmas that will yield the asymptotic normality of
$\sqrt{nh_{n}}\left(  \hat{\tau}_{n}-\tau_{n}\right)  ,$ and hence that of
$\sqrt{nh_{n}}\left(  \hat{\beta}_{n}-\bar{\beta}_{W}\right)  .$

Henceforth, to simplify notation we consider $m=1,$ although we indicate the
necessary changes in the arguments for $m>1.$ Define the function%
\begin{align*}
\zeta_{H}(Z,W)  &  =\alpha_{0H}(Z)+\alpha_{1H}(Z)W-\alpha_{0H}^{+}+\alpha
_{1H}^{+}W\\
&  -\left(  \dot{\alpha}_{0H}^{+}+\dot{\alpha}_{1H}^{+}W\right)  Z-\frac{1}%
{2}\left(  \ddot{\alpha}_{0H}^{+}+\ddot{\alpha}_{1H}^{+}W\right)  Z^{2},
\end{align*}
where $\dot{\alpha}_{0H}^{+}=\lim_{z\downarrow0}\partial\alpha_{0H}%
(z)/\partial z$ and $\ddot{\alpha}_{0H}^{+}=\lim_{z\downarrow0}\partial
^{2}\alpha_{0H}(z)/\partial z^{2},$ and similarly for $\alpha_{1H}.$ We use
later that
\[
\sup_{0<z<Mh_{n}}\left\vert \zeta_{H}(z,w)\right\vert =o(h_{n}^{2}%
)(1+\left\vert w\right\vert ).
\]
The function $\zeta_{H}(z,w)$ is the Taylor's remainder of order two of
$E[H_{i}|W_{i}=w,Z=z]:=\alpha_{0H}(z)+\alpha_{1H}(z)w\ $around $z=0.$ We can
also relate the coefficients in this expansion with the coefficients in
$\eta.$ Following the arguments above, it can be shown that
\[
\tilde{\eta}=\arg\min_{\eta}\sum_{i=1}^{n}\left(  H_{i}-\eta^{\prime}%
C_{i}\right)  ^{2}k_{h_{n}}(Z_{i}).
\]
estimates consistently $\eta.$ Thus,
\begin{align*}
\eta_{1}  &  =\alpha_{0H}(0),\text{ }\eta_{2}=\alpha_{1H}(0)\text{, }\eta
_{3}=\dot{\alpha}_{0H}(0),\\
\eta_{4}  &  =0,\text{ }\eta_{5}=\dot{\alpha}_{1H}(0),\text{ }\eta_{6}=0.
\end{align*}
Recall $U_{i}=Y_{i}-\tilde{X}_{i}^{\prime}\tau.$ Then, from the definitions
above%
\begin{align*}
E[U_{i}|W_{i},Z_{i}]  &  =E[H_{i}|W_{i},Z_{i}]-\alpha_{0H}^{+}+\alpha_{1H}%
^{+}W-\dot{\alpha}_{0H}^{+}Z+\dot{\alpha}_{1H}^{+}WZ\\
&  =\frac{1}{2}\left(  \ddot{\alpha}_{0H}^{+}+\ddot{\alpha}_{1H}^{+}%
W_{i}\right)  Z_{i}^{2}+\zeta_{H}(Z_{i},W_{i}).
\end{align*}
The following Lemmas make use of Assumptions \ref{assumption structural RDD},
\ref{Relevance} and \ref{AssumptionRRD}. Define $\mu_{\rho}(z)=E[W_{ij}^{\rho}|Z_{i}=z]$, for $\rho=1,2$, and their one-sided versions as usual.

\begin{lemma}
[Numerator: Expectation]\label{NumeratorExp}%
\[
E\left[  \frac{1}{nh_{n}}\sum_{i=1}^{n}S_{in}U_{i}k_{ih_{n}}\right]
\rightarrow_{p}\frac{1}{2}f(0)h_{n}^{2}(b_{U}+o(1)),
\]
where
\[
b_{U}=\left[
\begin{array}
[c]{c}%
\gamma_{2}(\ddot{\alpha}_{0H}^{+}+\ddot{\alpha}_{0H}^{+}\mu_{1}^{+})\\
\gamma_{3}(\ddot{\alpha}_{0H}^{+}+\ddot{\alpha}_{0H}^{+}\mu_{1}^{+})\\
\gamma_{2}(\ddot{\alpha}_{0H}^{+}\mu_{1}^{+}+\ddot{\alpha}_{0H}^{+}\mu_{2}%
^{+})\\
\gamma_{3}(\ddot{\alpha}_{0H}^{+}\mu_{1}^{+}+\ddot{\alpha}_{0H}^{+}\mu_{2}%
^{+})\\
\gamma_{2}(\ddot{\alpha}_{0H}^{-}+\ddot{\alpha}_{0H}^{+}\mu_{1}^{-})\\
\gamma_{3}(\ddot{\alpha}_{0H}^{-}+\ddot{\alpha}_{0H}^{-}\mu_{1}^{+})\\
\gamma_{2}(\ddot{\alpha}_{0H}^{-}\mu_{1}^{-}+\ddot{\alpha}_{0H}^{-}\mu_{2}%
^{+})\\
\gamma_{3}(\ddot{\alpha}_{0H}^{-}\mu_{1}^{+}+\ddot{\alpha}_{0H}^{-}\mu_{2}%
^{+})
\end{array}
\right]  .
\]

\begin{proof}
Let
\[
u_{l\rho}^{+}=\frac{1}{nh_{n}}\sum_{i=1}^{n}\left(  \frac{Z_{i}}{h_{n}%
}\right)  ^{l}W_{i}^{\rho}U_{i}k_{ih_{n}}^{+},\qquad l,\rho=0,1.
\]
Then, write%
\begin{align*}
E[u_{l\rho}^{+}]  &  =h_{n}^{-1}E\left[  \left(  \frac{Z_{i}}{h_{n}}\right)
^{l}W_{i}^{\rho}U_{i}k_{ih_{n}}^{+}\right] \\
&  =h_{n}^{-1}E\left[  \left(  \frac{Z_{i}}{h_{n}}\right)  ^{l}W_{i}^{\rho
}\left(  \frac{1}{2}\left(  \ddot{\alpha}_{0H}^{+}+\ddot{\alpha}_{1H}^{+}%
W_{i}\right)  Z_{i}^{2}+\zeta_{H}(Z_{i},W_{i})\right)  k_{ih_{n}}^{+}\right]
\\
&  =h_{n}^{-1}\frac{1}{2}\ddot{\alpha}_{0H}^{+}E\left[  \left(  \frac{Z_{i}%
}{h_{n}}\right)  ^{l}W_{i}^{\rho}Z_{i}^{2}k_{ih_{n}}^{+}\right]  +h_{n}%
^{-1}\frac{1}{2}\ddot{\alpha}_{1H}^{+}E\left[  \left(  \frac{Z_{i}}{h_{n}%
}\right)  ^{l}W_{i}^{\rho+1}Z_{i}^{2}k_{ih_{n}}^{+}\right] \\
&  +h_{n}^{-1}E\left[  \left(  \frac{Z_{i}}{h_{n}}\right)  ^{l}W_{i}^{\rho
}\zeta_{H}(Z_{i},W_{i})k_{ih_{n}}^{+}\right]  .
\end{align*}
By the change of variables $u=Z/h_{n},$%
\begin{align*}
h_{n}^{-1}E\left[  \left(  \frac{Z_{i}}{h_{n}}\right)  ^{l}W_{i}^{\rho}%
Z_{i}^{2}k_{ih_{n}}^{+}\right]   &  =h_{n}^{2}\int_{0}^{\infty}u^{l+2}%
k(u)\mu_{\rho}(uh_{n})f(uh_{n})du\\
&  =h_{n}^{2}\mu_{\rho}^{+}f(0^{+})\gamma_{l+2}+o(1),
\end{align*}
and similarly%
\[
h_{n}^{-1}E\left[  \left(  \frac{Z_{i}}{h_{n}}\right)  ^{l}W_{i}^{\rho+1}%
Z_{i}^{2}k_{ih_{n}}^{+}\right]  =h_{n}^{2}\mu_{\rho+1}^{+}f(0^{+})\gamma
_{l+2}+o(1).
\]
On the other hand, assume without loss of generality that $[-M,M]$ contains
the support of $k,$ so that
\[
h_{n}^{-1}E\left[  \left(  \frac{Z_{i}}{h_{n}}\right)  ^{l}W_{i}^{\rho}%
\zeta_{H}(Z_{i},W_{i})k_{ih_{n}}^{+}\right]  =o(h_{n}^{2}).
\]
The proof for the left limit version is analogous, and hence omitted.
\end{proof}
\end{lemma}

\begin{lemma}
[Numerator: Conditional Expectation]\label{NumeratorCondExp}%
\[
\frac{1}{nh_{n}}\sum_{i=1}^{n}E[S_{in}U_{i}k_{ih_{n}}|Z_{i},W_{i}]=\frac
{1}{nh_{n}}\sum_{i=1}^{n}E[S_{in}U_{i}k_{ih_{n}}]+o_{p}(h_{n}^{2}).
\]

\begin{proof}
We have
\begin{align*}
\frac{1}{nh_{n}}\sum_{i=1}^{n}E[S_{in}U_{i}k_{ih_{n}}|Z_{i},W_{i}]  &
=\frac{1}{nh_{n}}\sum_{i=1}^{n}S_{in}k_{ih_{n}}\left(  \frac{1}{2}\left(
\ddot{\alpha}_{0H}^{+}+\ddot{\alpha}_{1H}^{+}W_{i}\right)  Z_{i}^{2}+\zeta
_{H}(Z_{i},W_{i})\right) \\
&  =\frac{1}{2}\ddot{\alpha}_{0H}^{+}\frac{1}{nh_{n}}\sum_{i=1}^{n}%
S_{in}k_{ih_{n}}Z_{i}^{2}+\frac{1}{2}\ddot{\alpha}_{1H}^{+}\frac{1}{nh_{n}%
}\sum_{i=1}^{n}S_{in}k_{ih_{n}}W_{i}Z_{i}^{2}\\
&  +\frac{1}{nh_{n}}\sum_{i=1}^{n}S_{in}k_{ih_{n}}\zeta_{H}(Z_{i},W_{i}).
\end{align*}
Observe that
\begin{align*}
Var\left(  \frac{1}{nh_{n}}\sum_{i=1}^{n}S_{in}k_{ih_{n}}Z_{i}^{2}\right)   &
=\left(  nh_{n}^{2}\right)  ^{-1}Var\left(  S_{in}k_{ih_{n}}Z_{i}^{2}\right)
\\
&  \leq C\left(  nh_{n}\right)  ^{-1}h_{n}^{-1}E\left[  S_{in}S_{in}^{\prime
}k_{ih_{n}}^{2}Z_{i}^{4}\right] \\
&  =O\left(  \left(  nh_{n}\right)  ^{-1}h_{n}^{4}\right) \\
&  =o(1),
\end{align*}
since for $l,\rho=0,1,2$
\begin{align*}
h_{n}^{-1}E\left[  \left(  \frac{Z_{i}}{h_{n}}\right)  ^{l}W_{i}^{\rho
}k_{ih_{n}}^{+2}Z_{i}^{4}\right]   &  =h_{n}^{4}\int_{0}^{\infty}u^{l}%
k^{2}(u)\mu_{\rho}(uh_{n})f(uh_{n})du\\
&  =h_{n}^{4}\mu_{\rho}^{+}f(0^{+})v_{l}+o(1),
\end{align*}
where
\[
v_{l}=\int_{0}^{\infty}u^{l}k^{2}(u)du,
\]
and similarly%
\[
h_{n}^{-1}E\left[  \left(  \frac{Z_{i}}{h_{n}}\right)  ^{l}W_{i}^{\rho
}k_{ih_{n}}^{-2}Z_{i}^{4}\right]  =h_{n}^{4}\mu_{\rho}^{-}f(0^{-})v_{l}+o(1).
\]
Likewise,
\[
Var\left(  \frac{1}{nh_{n}}\sum_{i=1}^{n}S_{in}k_{ih_{n}}W_{i}Z_{i}%
^{2}\right)  =o(1).
\]
and
\[
Var\left(  \frac{1}{nh_{n}}\sum_{i=1}^{n}S_{in}k_{ih_{n}}\zeta_{H}(Z_{i}%
,W_{i})\right)  =o(1).
\]
Note that
\[
\frac{1}{nh_{n}}\sum_{i=1}^{n}S_{in}k_{ih_{n}}^{+}\left(  U_{i}-E[U_{i}%
|Z_{i},W_{i}]\right)  =\frac{1}{nh_{n}}\sum_{i=1}^{n}S_{in}k_{ih_{n}}^{+}%
\zeta_{i},
\]
where $\zeta_{i}=U_{i}-E[U_{i}|Z_{i},W_{i}]=\varepsilon_{Y_{i}}-\bar{\beta
}_{W}\varepsilon_{X_{i}}$ denotes the regression error. Then, we have the
following result.
\end{proof}
\end{lemma}

\begin{lemma}
[Numerator: Conditional Variance]\label{NumeratorConVar}%
\[
Var\left(  \frac{1}{nh_{n}}\sum_{i=1}^{n}S_{+in}k_{ih_{n}}\zeta_{i}\right)
=\frac{1}{nh_{n}}\Sigma_{U^{+}}+o(1),
\]
where
\[
\Sigma_{U^{+}}=f(0^{+})\left[
\begin{array}
[c]{cccc}%
v_{0} & v_{2} & v_{0}\sigma_{1}^{2+} & v_{2}\sigma_{1}^{2+}\\
v_{2} & v_{4} & v_{2}\sigma_{1}^{2+} & v_{2}\sigma_{2}^{2+}\\
v_{0}\sigma_{1}^{2+\prime} & v_{2}\sigma_{1}^{2+\prime} & v_{0}diag(\sigma
_{2}^{2+}) & v_{4}\sigma_{1}^{2+\prime}\\
v_{2}\sigma_{1}^{2+\prime} & v_{2}\sigma_{2}^{2+\prime} & v_{4}\sigma
_{1}^{2+\prime} & v_{4}diag(\sigma_{2}^{2+})
\end{array}
\right]  ,
\]%
\[
v_{l}=\int_{0}^{\infty}u^{l}k^{2}(u)du,\text{ }\sigma_{j\rho}^{2+}%
=\lim_{z\downarrow0}E[W_{ji}^{2\rho}\zeta_{i}^{2}|Z=z]\text{ and }\sigma
_{\rho}^{2+}=(\sigma_{1\rho}^{2+},...,\sigma_{c\rho}^{2+}).
\]

\begin{proof}
Consider the generic term, for $l,\rho=0,1,j=1,...,c,$%
\[
\frac{1}{nh_{n}}\sum_{i=1}^{n}\left(  \frac{Z_{i}}{h_{n}}\right)  ^{l}%
W_{ji}^{\rho}k_{ih_{n}}^{+}\zeta_{i},
\]
and its variance, which equals
\begin{align*}
\left(  nh_{n}\right)  ^{-1}h_{n}^{-1}E\left[  \left(  \frac{Z_{i}}{h_{n}%
}\right)  ^{2l}W_{ji}^{2\rho}k_{ih_{n}}^{+2}\zeta_{i}^{2}\right]   &  =\left(
nh_{n}\right)  ^{-1}\int_{0}^{\infty}u^{2l}k^{2}(u)\sigma_{j\rho}^{2}%
(uh_{n})f(uh_{n})du\\
&  =\left(  nh_{n}\right)  ^{-1}f(0^{+})\sigma_{j\rho}^{2+}v_{2l}+o(1)
\end{align*}
where $\sigma_{j\rho}^{2}(z)=E[W_{ji}^{2\rho}\zeta_{i}^{2}|Z_{i}=z].$
Similarly, we have the following result, which proof is the same as in the
previous lemma.
\end{proof}
\end{lemma}

\begin{lemma}
\label{NumeratorConVarLeft}%
\[
Var\left(  \frac{1}{nh_{n}}\sum_{i=1}^{n}S_{-in}k_{ih_{n}}\zeta_{i}\right)
=\frac{1}{nh_{n}}\Sigma_{U^{-}}+o(1),
\]
where
\[
\Sigma_{U^{-}}=f(0^{-})\left[
\begin{array}
[c]{cccc}%
v_{0} & v_{2} & v_{0}\sigma_{1}^{2-} & v_{2}\sigma_{1}^{2-}\\
v_{2} & v_{4} & v_{2}\sigma_{1}^{2-} & v_{2}\sigma_{2}^{2-}\\
v_{0}\sigma_{1}^{2-\prime} & v_{2}\sigma_{1}^{2-\prime} & v_{0}diag(\sigma
_{2}^{2-}) & v_{4}\sigma_{1}^{2-\prime}\\
v_{2}\sigma_{1}^{2-\prime} & v_{2}\sigma_{2}^{2-\prime} & v_{4}\sigma
_{1}^{2-\prime} & v_{4}diag(\sigma_{2}^{2-})
\end{array}
\right]  ,
\]%
\[
\sigma_{j\rho}^{2-}=\lim_{z\downarrow0}E[W_{ji}^{2\rho}\zeta_{i}%
^{2}|Z=z]\text{ and }\sigma_{\rho}^{2-}=(\sigma_{1\rho}^{2-},...,\sigma
_{c\rho}^{2-}).
\]

\end{lemma}

Define%
\[
\Sigma_{U}=\left[
\begin{array}
[c]{cc}%
\Sigma_{U^{+}} & 0\\
0 & \Sigma_{U^{-}}%
\end{array}
\right]
\]

\begin{lemma}
[Numerator: Conditional CLT]\label{NumeratorCondCLT}%
\[
(nh_{n})^{-1/2}\sum_{i=1}^{n}S_{in}k_{ih_{n}}\zeta_{i}\rightarrow_{d}N\left(
0,\Sigma_{U}\right)  .
\]

\begin{proof}
Consider a generic term for $l,\rho=0,1,j=1,...,c,$%
\[
\frac{1}{\sqrt{nh_{n}}}\sum_{i=1}^{n}\left(  \frac{Z_{i}}{h_{n}}\right)
^{l}W_{ji}^{\rho}k_{ih_{n}}^{+}\zeta_{i}.
\]
We apply Lyapounov with third absolute moment. By the lemma on the asymptotic
variance, we need to establish%
\[
\left(  nh_{n}\right)  ^{-1/2}h_{n}^{-1}E\left[  \left(  \frac{Z_{i}}{h_{n}%
}\right)  ^{3l}W_{ji}^{3\rho}k_{ih_{n}}^{+3}\zeta_{i}^{3}\right]  =o(1).
\]
But note that, defining $\kappa_{j\rho}(z)=E[W_{ji}^{3\rho}\zeta_{i}%
^{3}|Z=z],$
\begin{align*}
h_{n}^{-1}E\left[  \left(  \frac{Z_{i}}{h_{n}}\right)  ^{3l}W_{ji}^{3\rho
}k_{ih_{n}}^{+3}\zeta_{i}^{3}\right]   &  =\int_{0}^{\infty}u^{3l}%
k^{3}(u)\kappa_{j\rho}(uh_{n})f(uh_{n})du\\
&  =O(1).
\end{align*}
The same holds for the left limit part.
\end{proof}
\end{lemma}

\begin{lemma}
[Numerator: Unconditional CLT]\label{NumeratorUncCLT}%
\[
(nh_{n})^{-1/2}\sum_{i=1}^{n}S_{in}U_{i}k_{ih_{n}}-\frac{(nh_{n})^{1/2}%
h_{n}^{2}}{2}f(0)b_{U}\rightarrow_{d}N\left(  0,\Sigma_{U}\right)  .
\]

\begin{proof}
It follows from previous Lemmas.
\end{proof}
\end{lemma}

\begin{lemma}
[Main CLT]\label{MainCLT}%
\[
\sqrt{nh_{n}}(\hat{\tau}_{n}-\tau_{n})\rightarrow_{d}N\left(  0,\Omega\right)
,
\]
where
\[
\Omega=\left(  \Delta\Gamma^{-1}\Delta^{\prime}\right)  ^{-1}\Delta\Gamma
^{-1}\Sigma_{U}\Gamma^{-1}\Delta^{\prime}\left(  \Delta\Gamma^{-1}%
\Delta^{\prime}\right)  ^{-1}.
\]

\begin{proof}
It follows from previous Lemmas.
\end{proof}
\end{lemma}

\end{document}